\documentclass[pre,twocolumn,superscriptaddress,showpacs,floatfix]{revtex4-1}
\usepackage{graphicx}
\usepackage{amssymb,amsmath,amsbsy,amsfonts,bm}
\usepackage[percent]{overpic}
\usepackage{color}

\begin{document}
\title{On the yielding of a defect-rich model crystal under shear: \\ insights from molecular dynamics simulations}

 \author{Gaurav P. Shrivastav}
 \affiliation{Institut f\"ur Theoretische Physik and
   Center for Computational Materials Science (CMS), TU Wien,
   Wiedner~Hauptstra{\ss}e 8-10, A-1040 Wien, Austria}
 \author{Gerhard Kahl}
 \affiliation{Institut f\"ur Theoretische Physik and
   Center for Computational Materials Science (CMS), TU Wien,
   Wiedner~Hauptstra{\ss}e 8-10, A-1040 Wien, Austria}
\date{\today}

\begin{abstract}
Point defects in real crystals at finite temperatures are inevitable. Their dynamics severely influence the mechanical properties of crystals under shear giving rise to nonlinear effects such as ductility. Therefore, it is crucial to explore the interplay of the equilibrium point-defect diffusion timescales and shear-induced timescales to understand the plastic deformation of crystals. Using extensive nonequilibrium molecular dynamics simulations, we present a study on the yielding behavior of cluster crystals (CC), an archetypal model for defect-rich crystals where the crystalline structure is characterized by multiple occupancies (cluster) of particles at a lattice site. In equilibrium, particles diffuse via site-to-site hopping keeping the crystalline structure intact. We consider the CCs at a fixed density and different temperatures where it remains in the FCC structure, and the diffusion timescales of particles vary depending on the temperature. We choose the range of shear rates, which covers timescales higher and much lower than the equilibrium diffusion timescales at high temperatures. For the considered range of shear rates and temperatures, both the macro and microscopic responses of CCs to the shear suggest that the yielding scenario remains independent of the diffusion of particles. It involves the plastic deformation of the underlying crystalline structure. The averaged local bond order parameters and local angle measurements demonstrate the structural changes and cooperative movement of the center of masses of the clusters close to the yield point. A comparison with the soft-sphere (SS) FCC crystal reveals the similarities in the yielding behavior of both systems. Nonetheless, diffusion of particles influences certain features, such as a less prominent increase in the local bond order parameters and local angles close to the yield point. Our simulations provide an insight into the role of the diffusion of defects in the yielding behavior of cluster crystals under shear, and these will be helpful in the development of theories for the plastic deformation of defect-rich crystals.
\end{abstract}
\maketitle

\section{Introduction}
\label{sec:introduction}

Defects play a crucial role in determining the mechanical properties of crystals \cite{sethna2017deformation}. Real crystals at finite temperatures contain various kinds of defects, such as dislocations, vacancies, interstitials, etc. Among these, the role of dislocations (line defects) in modifying the mechanical response of crystals has been well understood in theory and experiments \cite{friedel2013dislocations,schall2006visualizing}. Large scale molecular dynamics (MD) simulations have established the connection between interatomic processes and mesoscopic behavior predicted by {\it dislocation-dynamics} simulations \cite{bulatov1998connecting}. However, the role of point defects such as interstitials and vacancies on the mechanical properties of crystals is rather less explored. Continuum elasticity theory considers ``ideal crystals'' with no point defects and explains the rigidity of crystals as a result of spontaneous breaking of continuous translational symmetry \cite{chaikin1995principles}. Recently proposed microscopic theory \cite{walz2010displacement}, developed under the framework of linear-response and correlation functions theory, incorporates point defects and successfully predicts elastic constants, point-defect density, dispersion relations, etc. \cite{haring2015coarse} for non-ideal crystals. However, a further extension is needed to understand the yielding behavior and other non-linear effects, such as the transition from brittle to ductile mechanical response, which are often reported in experiments \cite{cuitino1996ductile,noell2018mechanisms}. 

The response of ideal crystals to the deformation is markedly different from non-Newtonian liquids or amorphous solids. The former responds to the shear via the formation of slip planes, which allow full release of stresses \cite{nath2018existence} while the latter responds via the diffusion of particles or formation of inhomogeneous flow patterns, known as shear bands -regions where stresses localize \cite{bonn2017yield,golkia2020flow,shrivastav2016yielding}. The yielding of ideal crystals and amorphous solids can be viewed as a timescale phenomenon with the difference that in the case of crystals the viscosity diverges as the shear stress tends to zero \cite{sausset2010solids,nath2018existence}. For point defect-rich crystals, one expects that the diffusion of point-defects should alter the yielding behavior and allow crystals to flow easily. It should be mentioned here that in viscoelastic materials where diffusion of particles is the primary mechanism for structural relaxation, the Newtonian response occurs when shear-induced timescales are comparable with the diffusion time scales \cite{zausch2008equilibrium,golkia2020flow}. Therefore, it is necessary to understand the interplay of shear-induced and the point-defect diffusion timescales for defect-rich crystals under shear. For crystals with point defects, these effects are hitherto not explored in computer simulations. The reason may be a low concentration \cite{pronk2001point} and long-ranged interaction of point defects due to strain fields \cite{van2017diffusion}, which make their collective dynamics slower at high densities. To this end, one requires a suitable model system with a large density and faster collective dynamics of point defects. 

We consider {\it cluster crystals}, a soft-matter system of particles interacting via purely repulsive ultrasoft potential known as the generalized exponential model with index $n$ (GEM-n) \cite{mladek2006formation,mladek2007clustering}. This interaction allows a complete overlap of particles with a {\it finite} energy penalty at vanishing interparticle distance. This leads to cluster phases, which are possible if the Fourier transform of the potentials shows negative components \cite{likos2001criterion}. These systems show crystalline phases, BCC and FCC, at sufficiently low temperatures and high densities that are characterized by a monodisperse cluster of particles occupying lattice sites \cite{mladek2006formation,zhang2010reentrant,wilding2013monte,mladek2007clustering, likos2008cluster,mladek2008multiple}. 
Although the clusters are localized at lattice sites, the arrangement of particles inside clusters remains disordered. In equilibrium, particles vibrate inside clusters, and at high temperatures hop from one cluster to the other at the neighboring lattice site. The longtime dynamics is diffusive \cite{coslovich2011hopping} and the distribution of the jump length is found to be exponential at short distances. At large distances, this distribution follows a power-law decay, i.e., a behavior reminiscent of L\'{e}vy flights \cite{coslovich2011hopping}. Recent {\it out-of-equilibrium} investigations explore the steady-state behavior of cluster crystals under shear and find that these systems show shear-induced fluidization and string formation at high shear rates \cite{nikoubashman2011cluster,nikoubashman2012flow,nikoubashman2013computer}. However, yielding behavior remains to be far from being understood. From an experimental point of view, colloids have provided a potential platform to test theoretical predictions in-and-out of equilibrium \cite{habdas2002video,lowen1994melting}. Cluster crystals are also easily amenable to experiments as monomer resolved computer simulations predict that suitably synthesized amphiphilic dendrimers of the second generation can be used as penetrable soft colloids \cite{lenz2011monomer,lenz2012microscopically}.

Cluster crystals can be considered as an archetypal model system for defect-rich crystals as fluctuating number density of particles at lattice sites corresponds to interstitials. Furthermore, particles can hop from one cluster to the nearby cluster creating a vacancy at the parent lattice site. Of course, this vacancy will be occupied by particle hopping from other lattice sites. In our previous work \cite{shrivastav2020on}, we have demonstrated that cluster crystals exhibit an overshoot in their stress-strain response. For the fixed temperature, the height of this overshoot decreases with a decreasing shear rate. The range of temperatures considered there was limited to high temperatures where long time dynamics in equilibrium is always found to be diffusive. In this work, we consider a broader range of temperatures and, using extensive non-equilibrium MD simulations, study the changes in the structure and dynamics of cluster crystals under shear. The work aims to understand the interplay of defect-diffusion and shear-induced timescales. Hence, we shall focus on a sufficiently high temperature where the equilibrium mean-square displacement (MSD) of particle shows the longtime diffusive behavior. The range of the shear rates is considered such that it covers a wide range of timescales, large and small, compared to the equilibrium diffusion timescale. Our MSD data under shear and further analyses with the center of mass (COM) of clusters will reveal that in cluster crystals, diffusion of particles is not the primary mechanism for stress relaxation. Instead, it is the plastic deformation of the underlying FCC structure, henceforth {\it skeleton}, which governs the yielding behavior. Further comparison with an FCC crystal formed by particles interacting via purely repulsive soft-sphere potential will reveal that defects modify the transient response of cluster crystals. However, overall yielding behavior remains independent of temperature and diffusion of particles.

The rest of the paper is organized as follows: in Section~\ref{sec:model_simulations}, we introduce the model and give details of the simulations and the related protocols. Results are presented and discussed in Section~\ref{sec:results}, while the final section contains the summary of results, concluding remarks, and an outlook to future, related investigations.

\section{System and simulation methods}
\label{sec:model_simulations}

\subsection{Simulation of the cluster crystal}
\label{subsec:sim_cluster}

In our cluster crystal system particles interact via the generalized exponential (GEM-$n$) potential \cite{mladek2006formation}. Similar as in preceding contributions \cite{coslovich2011hopping,mladek2006formation} we set $n = 4$, thus the interaction potential is given by
\begin{eqnarray}
\label{gem4}
\Phi^{\rm GEM}(r) = \epsilon \exp\left[-\left(r/d \right)^{4}\right] ;
\end{eqnarray}
here $r$ is the distance between two particles, $d$ and $\epsilon$ set the length- and energy-scales of the model, respectively; in contrast to the usual notation we use the symbol $d$ for the range of the interaction, since the conventional symbol $\sigma$ is reserved to denote in this manuscript the stress. We truncate the potential at a distance $r_{\rm c} = 2.2d$; $\Phi^{\rm GEM}(r)$ is then shifted so that it smoothly vanishes at $r_{\rm c}$. Temperature $T$, density $\rho$, and time $t$, are measured in the units of $k_{\rm B}T/\epsilon$, $\rho d^{3}$, and $t_0 = d\sqrt{m/\epsilon}$, respectively; further, $m$ is the mass of particles and $k_{\rm B}$ is the Boltzmann constant. In the following we set the values of $\epsilon$, $d$, $m$, and $k_{\rm B}$ equal to unity.

We perform non-equilibrium molecular dynamics (MD) simulations in an NVT-ensemble where the number of particles, $N$, the volume, $V$, and the temperature, $T$, of the system are fixed. All  simulations are  carried out using the LAMMPS package \cite{plimpton1995fast}. In this contribution we consider ensembles with $N = 416, 1300, 3328, 6500, 26624$, and $52000$ particles; throughout we use a fixed density $\rho = 6.5$ while for the temperature a set of values, namely $T = 0.4, 0.5, 0.6, 0.7$ has been considered: from literature it is known that at this density and at these temperatures the system is in a stable FCC cluster phase, where each site of the FCC lattice is occupied by a cluster of overlapping particles (see, for instance, the phase diagram shown in Ref. \cite{mladek2006formation}). Data available in the literature \cite{mladek2006formation,coslovich2011hopping,mladek2007clustering} provide evidence that the average number of particles pertaining to a cluster, $N_{\rm c}$, assumes for the considered state points a value $N_{\rm c} \simeq 13$ and lattice constant $l_{a} = 2$. Most of our calculations are based on ensembles of 3328 and 26624 particles, corresponding thus to systems with 256 and 2048 clusters, respectively.

The temperature of the system is fixed via a thermostat, using dissipative particle dynamics (DPD) \cite{soddemann2003dissipative}. The DPD equation-of-motions read (where the dot represents the time-derivative of the respective quantity): 
\begin{eqnarray}
\label{neq}
\dot{\bm r}_{i} &=& \frac{{\bm p}_{i}}{m_{i}}, \\
\label{dpdeq}
\dot{\bm p}_{i} &=& \sum_{j\neq i} \left[{\bm F}_{ij} 
+ {\bm F}^{\rm D}_{ij} + {\bm F}^{\rm R}_{ij}\right], ~~~~ i = 1, \dots, N .
\end{eqnarray}
${\bm r}_{i}$ is the position and ${\bm p}_{i}$ is the momentum of the particle with index $i$. The conservative force, ${\bm F}_{ij}$, acting on a pair of particles $i$ and $j$ can be readily calculated from  the interparticle interaction defined in Eq.~(\ref{gem4}). The dissipative force, ${\bm F}^{\rm
D}_{ij}$, is given by
\begin{eqnarray}
\label{dissf}
{\bm F}^{\rm D}_{ij} = -\zeta \omega^{2}\left(r_{ij}\right) 
\left(\hat{\bm r}_{ij}\cdot {\bm v}_{ij}\right)\hat{\bm r}_{ij} ; 
\end{eqnarray}
${\bm r}_{ij}$ the distance vector between particles $i$ and $j$, $\hat{\bm r}_{ij}$ is the unit vector of ${\bm r}_{ij}$, and $r_{ij}$ the distance between the two particles; further, ${\bm v}_{ij} = ({\bm v}_{i} - {\bm v}_{j})$ is the relative velocity between the particles $i$ and $j$, and $\zeta$ is the friction coefficient; the value of $\zeta$ is set to unity. Furthermore, $\omega\left(r_{ij}\right)$ is a distance-dependent weight function which defines the range of interaction for the dissipative and random forces. In order to associate the continuous stochastic differential equation with the DPD algorithm \cite{hoogerbrugge1992simulating} the usual choice for $\omega\left(r_{ij}\right)$ is as follows \cite{espanol1995statistical}:
\begin{eqnarray}
\label{omega}
\omega (r_{ij}) = 
\begin{cases}
1-r_{ij}/R_{c} & \quad {\rm if} ~~ 0 \le r_{ij} \le R_{\rm c},\\
0                     & \quad {\rm otherwise}.
\end{cases}
\end{eqnarray}
For the cutoff radius of this function, $R_{\rm c}$, we have taken for simplicity the same value as for $r_{\rm c}$, i.e., $R_{\rm c} = r_{\rm c} = 2.2d$. 

Eventually, the ${\bm F}^{\rm R}_{ij}$  represent in Eq.~(\ref{dpdeq}) random forces, defined as
\begin{eqnarray}
\label{ranf}
F^{\rm R}_{ij} = 
\sqrt{2k_{B}T\zeta}\omega (r_{ij})\theta_{ij}\hat{\bm r}_{ij}\, .
\end{eqnarray}
Here, the $\theta_{ij}$ are uniformly-distributed random numbers with zero mean and unit variance. For further details about the parameters of the DPD thermostat we refer to  Refs.~\cite{zausch2008equilibrium,golkia2020flow}. 

The equations-of-motion, i.e., Eqs.~(\ref{neq}) and (\ref{dpdeq}), are integrated via the velocity Verlet algorithm using an integration time step $\Delta t = 0.005 t_0$ \cite{allen2017computer}. 

The initial configurations for our simulations are ideal FCC cluster crystals where each lattice site is occupied by $N_{\rm c} = 13$ completely overlapping particles and assuming a lattice constant that is compatible with the chosen value of the density, i.e. $\rho = 6.5$. Starting from this configuration, the system is equilibrated over $10^{6}$ MD steps at a temperature $T = 0.8$. The now equilibrated system is further evolved over $5\times 10^{6}$ MD steps (where it has reached the diffusive regime), storing on a regular basis configurations in intervals of $10^{5}$ MD steps. These stored configurations then serve as independent initial configurations for subsequent simulations: from each of these state points, 50 independent simulation runs have been launched. Particle configurations of the system have been stored during the run at logarithmic time intervals; observables were then obtained by averaging the related quantity over these runs.

We impose planar Couette flow on the bulk cluster crystal via Lees-Edwards boundary conditions \cite{lees1972computer}. The shear is applied in the $(x, z)$-plane along the $x$-direction; thus, the $z$- and $y$-directions are the gradient and vorticity directions, respectively, while $x$ is the shear-direction. In this study the range of the shear rates, $\dot \gamma$, extends from $\dot{\gamma} = 10^{-7}$ to $\dot \gamma = 10^{-1}$ (in units of $t_0^{-1}$).  

\subsection{Simulations on the soft sphere crystal}
\label{subsec:sim_hard}

For the soft-sphere (SS) FCC crystals we assumed that the particles interact via a Weeks-Chandler-Andersen (WCA) potential, being defined as \cite{weeks1971role}:
\begin{eqnarray}
\label{wca}
\Phi^{\rm HS} (r) = 
\begin{cases}
4 \epsilon \left[
\left(\frac{d}{r}\right)^{12}- \left(\frac{d}{r}\right)^{6} +  \frac{1}{4} \right]               & \quad {\rm if} ~~ 0 \le r_{ij} \le r^{\rm SS}_{\rm c},\\
0                     & \quad {\rm otherwise} ;
\end{cases}
\end{eqnarray}
$r$ is the distance between two particles, $d$ and $\epsilon$ are again length- and energy-scale of the system and are both set to unity. The potential is truncated at a distance $r^{\rm SS}_{c} = 2^{1/6}$ d; thus this shifted potential vanishes smoothly at $r^{\rm SS}_{c}$.   For the SS FCC crystal we consider $N = 4000$ particles; for the temperature and the density we assume the following values: $T = 0.01$ and $\rho = 1.2$. Starting from an ideal FCC crystal as an initial configuration we equilibrate the system over $1\times 10^{7}$ MD steps using again a DPD thermostat (with the same parameters as in the cluster crystal case). We further shear this system with a shear rate $\dot{\gamma} = 10^{-6}$ using the same protocol as for the cluster crystals.

\section{Results}
\label{sec:results}

\subsection{The system in equilibrium}
\label{subsec:equib}

\begin{figure*}
\centerline{\includegraphics[width=0.8\textwidth]{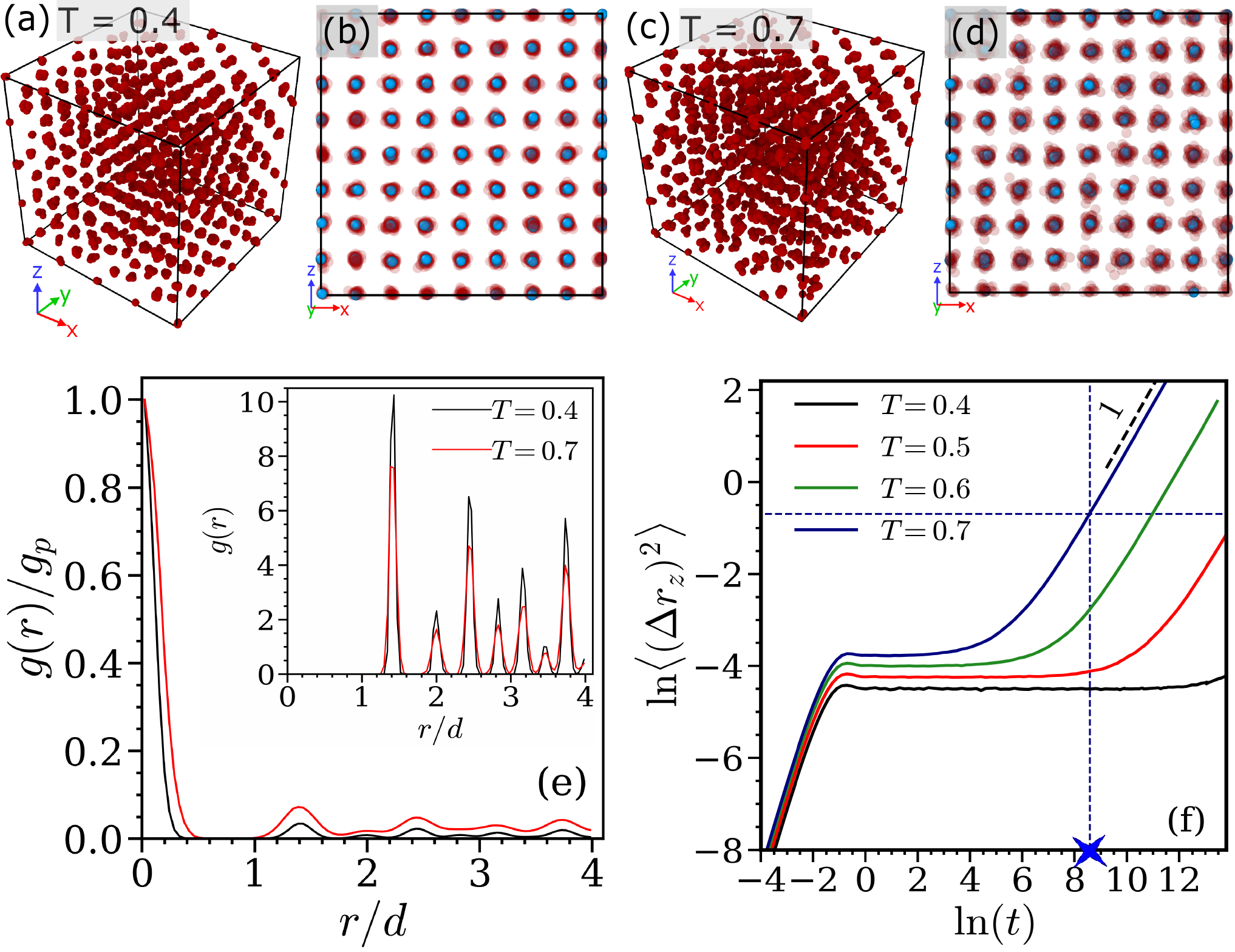}}
\caption{Panels (a) and (d): cluster crystal configurations at equilibrium at temperatures $T = 0.4$ and $T = 0.7$, respectively. Panels (b) and (d): front view of the snapshots shown in panels (a) and (c). Blue spheres represent the centers of mass of clusters. Panel (e): pair correlation function, $g(r)$, of the clusters as a function of distance, $r$, at temperatures $T = 0.4$ and $T = 0.7$ (as labeled in the inset). The $g(r)$ are scaled for different temperatures by the value of the first peak, $g_{\rm p}$, in the pair correlation functions. The inset shows $g(r)$ of the centers of mass of the clusters; these positions are highlighted in panels (b) and (d) as black spheres. Panel (f): $z$-component of the mean-square displacement (MSD) of ultrasoft particles, calculated  at different temperatures (as labeled). The horizontal, blue-dashed line indicates where the MSD assumes a value of $\sigma^2/2$; from the intersection point of this line with the MSD-curve for $T = 0.7$ a vertical, dashed-blue line projects down to the related hopping timescale, $\tau_{\rm h} ( T = 0.7)$ (see text): $\tau_{h} (T = 0.7) = 5349.44$ is marked by blue cross on the time axis. The black dashed line represents a line with slope $1$, indicating a diffusive behaviour.}
\label{fig1}
\end{figure*}

In a first step we consider the system in its equilibrium. To this end we briefly summarize the self-assembly and the dynamics of cluster crystals in equilibrium. Panels (a) and (b) of Fig.~\ref{fig1} show simulation generated equilibrium configurations of cluster crystals at a density $\rho = 6.5$ for two different temperatures, namely $T = 0.4$ and $T = 0.7$. The red, semi-transparent, overlapping spheres show the clusters, while the center of mass (COM) of each cluster is plotted as a black sphere. In panel (c) of  Fig.~\ref{fig1} we plot the pair-correlation function, $g(r)$, of the COM of the clusters as a function of distance, $r$, for $T = 0.4$ and $T = 0.7$. The $g(r)$ display distinct peaks at distances $r \sim 1.41\sigma$ and $r \sim 2\sigma$ which are the nearest- and the second-nearest neighbor distances in an FCC crystal with a lattice constant equal to $2\sigma$. Of course the the peaks in the $g(r)$ broaden at high temperatures indicating the enhanced fluctuations in the COM positions of the clusters. 

The $z$-component of the mean-square displacement (MSD) of the individual particles, $\langle (\Delta r_z)^2 \rangle$, is shown for the equilibrium states at temperatures $T = 0.4, 0.6$, and $0.7$ in panel (d) of Fig.~\ref{fig1}. We only display the $z$-component of the MSD since in our subsequent shear simulations the $z$-axis denotes the gradient direction; thus, $\langle (\Delta r_z)^2 \rangle$ quantifies the non-affine displacement of the particles under shear. Similar to a previous study \cite{coslovich2011hopping}, we observe that the MSDs display for all temperatures investigated a short-time, ballistic regime which at later times levels off to a plateau region. At the lowest temperature, i.e., at $T = 0.4$, this plateau persists over the entire simulation time window. In contrast, at the highest temperature investigated, i.e., at $T = 0.7$, the MSD shows at larger times a cross-over to a now diffusive regime: this feature originates from hopping processes of individual particles, migrating from one cluster to a neighboring one. In this context it should be emphasized that -- despite these hopping events -- the underlying FCC structure of the clusters remains intact \cite{coslovich2011hopping}. The timescale of these hopping processes, $\tau_{\rm h}$, can be estimated by identifying the time where $\langle (\Delta r_z)^2 \rangle$ attains a  value of $\sigma^{2}/2$, corresponding to a distance of half of the nearest neighbour distance in an FCC lattice -- note that the nearest neighbor distance in our case is $\sqrt{2}\sigma$; this value of the MSD is highlighted in panel (d) of Fig.~\ref{fig1} by a horizontal, blue-dashed line. Intersecting this line with the MSD-curve leads -- by projection onto the time-axis -- to $\tau_{\rm h} = \tau_{\rm h} (T)$; for the temperature $T = 0.7$, $\tau_{\rm h}$ is highlighted in panel (d) of Fig.~\ref{fig1} by a blue cross. In the following we use the value of  $\tau_{\rm h}$ to classify {\it high} and {\it low} shear rates in our simulations: introducing the shear-induced timescale, $\tau_{s} \left(= 1/\dot{\gamma}\right)$ we consider shear rates with $\tau_{\rm s} < \tau_{\rm h}$ as high, while shear rates with $\tau_{\rm s} > \tau_{\rm h}$ are considered as low. 

In the subsequent shear simulations on cluster crystals we focus on two temperatures: (i) $T = 0.7$ for which $\tau_{\rm h}$ attains a value that is smaller than the total simulation time; therefore we can access for this temperature both the high and the low shear rate windows; (ii) at $T = 0.4$ the value of $\tau_{\rm h}$ is considerably larger than the accessible total simulation time; hence, at this temperature all shear rates are considered as high. 

\subsection{The system under shear}
\label{subsec:nonequb}

\subsubsection{Stress vs. strain response}
\label{subsubsec:str_vs_str}

We start our investigations of the non-equilibrium properties by measuring the stress of the system, $\sigma_{xz}(t)$, as a function of strain, $\dot{\gamma}t$, for various shear rates, $\dot \gamma$, and at different temperatures. We calculate $\sigma_{xz}$ via the Irving-Kirkwood expression  \cite{irving1950statistical}:
\begin{eqnarray}
\label{str}
\langle \sigma_{xz}(t)\rangle &=&
\frac{1}{V}\bigg\langle\sum_{i}\big[mv_{i,x}(t) v_{i,z}(t)                      
      + \sum_{i>j} r_{ij,x}(t) F_{ij,z}(t) \big]\bigg\rangle.
\end{eqnarray}
Here, $m$ is the mass of the particles, $v_{i,x}(t)$ and $v_{i,z}(t)$ represent the $x$- and $z$-components of the velocity of particle $i$; further, $r_{ij,x}(t)$ is the $x$-component of the displacement vector between particles $i$ and $j$, $F_{ij,z}(t)$ denotes the $z$-component of the force between particles $i$ and $j$, and $V$ represents the total volume of the system.  The angular brackets in Eq.~(\ref{str}) stand for an averaging procedure, that is carried out over 50 independent runs (see above). We note that the kinetic terms in Eq.~(\ref{str}), i.e. the one proportional to $m_{i}v_{i,x}v_{i,z}$, turn out to be very small; therefore we have neglected them for the calculation of the shear stress.

\begin{figure}
\centerline{\includegraphics[width=0.5\textwidth]{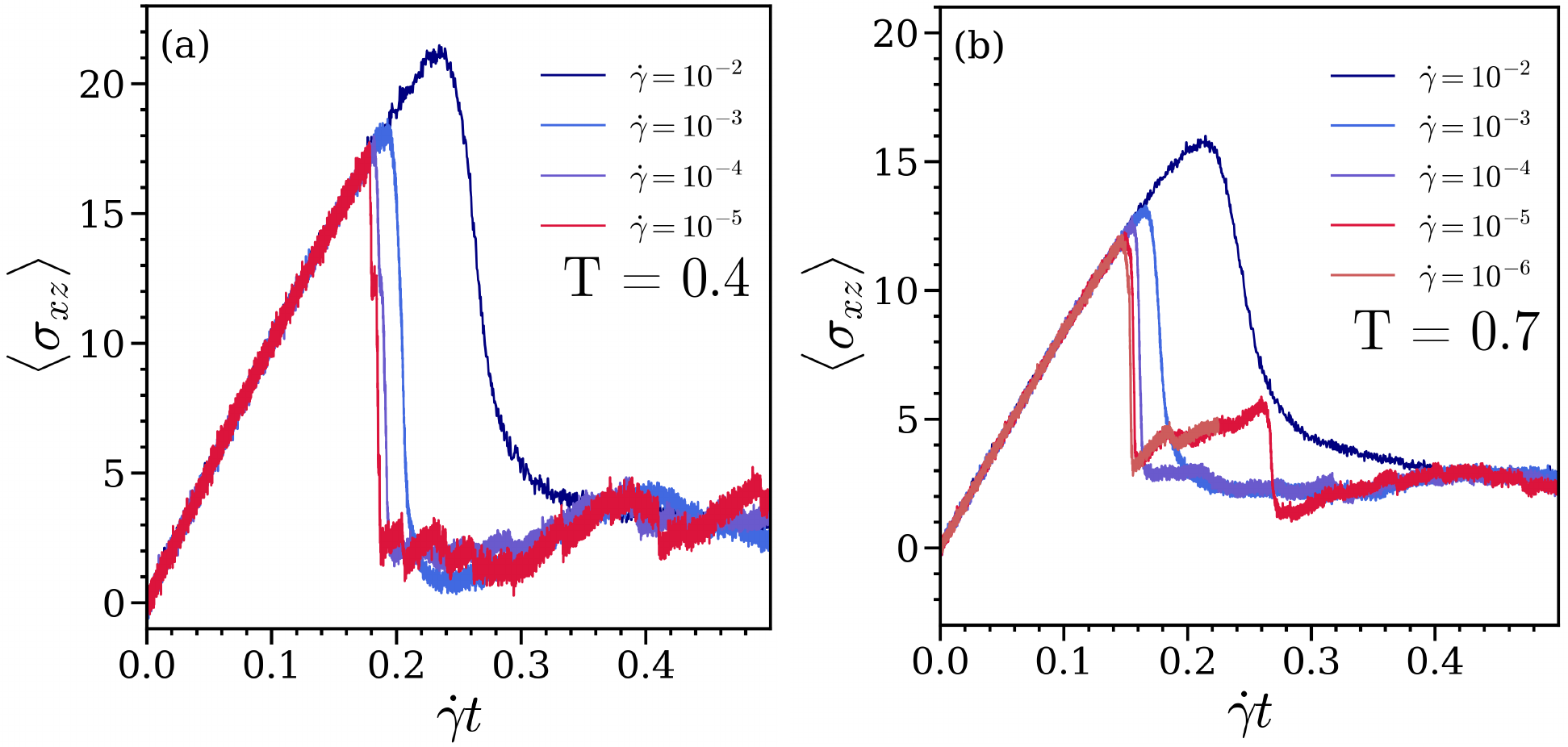}}
\caption{Panels (a) and (b): time evolution of the stress, $\langle\sigma_{xz} \rangle$, for the cluster crystal as a function of strain, $\dot{\gamma}t$, for different shear rates (as labeled), for the temperatures $T = 0.4$ (a) and $T = 0.7$ (b). Results are based on ensembles of $N = 26624$ particles.}
\label{fig2}
\end{figure}

\begin{figure}
\centerline{\includegraphics[width=0.5\textwidth]{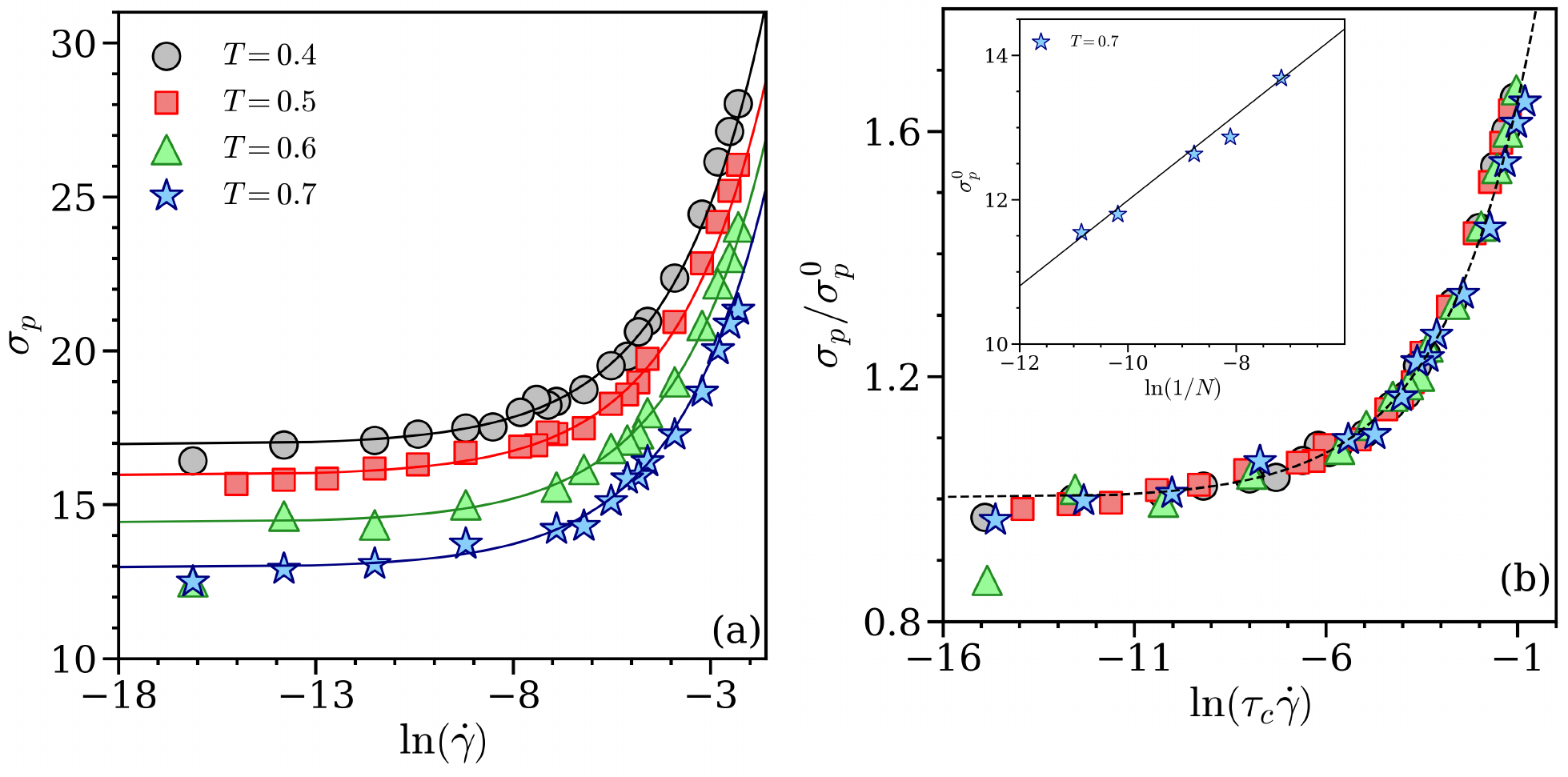}}
\caption{Panel (a): variation of the maximum of the stress-strain curves, $\sigma_{\rm p}$, as a function of the shear rate, $\dot{\gamma}$, for different temperatures (as labeled). In this panel, the solid lines represent the Herschel-Bulkley type function defined in Eq.~(\ref{hb}), using suitably fitted parameters $\sigma_{\rm p}^0$, $A$, and $\gamma$. Panel (b): scaled curves $\sigma_{\rm p}$ vs. shear rate (shown in panel (a), using Eq.~(\ref{sclhb})); all data collapse on a master curve. The black dashed line represents the scaling function given in Eq. (\ref{sclhb}). Results are based on ensembles of $N = 3328$ particles. The inset in panel (b) shows the variation of $\sigma_{\rm p}^{0}$ as function of the system size, considering ensembles of $N = 1300, 3328, 6500, 26624$ and 52000 particles  at $T = 0.7$. The black solid line represent the logarithmic fit given in Eq.~(\ref{logfit}) (see text).}
\label{fig2a}
\end{figure}

Panels (a) and (b) in Fig.~\ref{fig2} display the response of the stress on the strain for cluster crystals, calculated  at two different temperatures, namely $T = 0.4$ and $0.7$; in these calculations ensembles of 26624 particles were considered. The stress first increases, reaches a maximum and then drops suddenly; this feature corresponds to the yielding of the cluster. We observe that the drop in the stress becomes sharper and that the height of the stress maximum, $\sigma_{\rm p}$, increases as we  decreases the shear rate; thesse features are observed for all shear rates and temperatures considered. The data shown in panels (a) and (b) of Fig.~\ref{fig2} also provide evidence that for a given shear rate the peak of the stress-strain curve is more pronounced at low temperatures. 

This feature is further analysed via the data shown in panel (a) of Fig.~\ref{fig2a} where the maximum of the stress-strain curve is plotted as a function of shear rate for different temperatures on a much finer $\dot \gamma$-grid. We note in passing that these results are now -- as a tribute to the high computational costs -- based on simulations of ensembles of $N = 3328$ particles; a brief analysis of the data in terms of system size will be given below. High values of $\sigma_{\rm p}$, observed in particular at low temperatures indicate the higher rigidity of the cluster crystal; in contrast, at high temperatures the higher diffusivity of particles destabilizes this rigidity, leading to lower values of $\sigma_{\rm p}$. We also note that $\sigma_{\rm p}$ curves, calculated for different values of $\dot{\gamma}$ (as displayed in panel (a) of Fig.~\ref{fig2a}) can be mapped via a Herschel-Bulkley (HB) type function onto a single master curve; this HB function is given by \cite{l99,bonn2017yield}
\begin{eqnarray}
\label{hb}
\sigma_{\rm p}\left(T, \dot{\gamma}\right) =
\sigma^{0}_{\rm p}\left(T\right) + A(T) \dot{\gamma}^{\alpha} 
\end{eqnarray}
with adjustable parameters $\sigma_{\rm p}^{0}$, $A$, and $\alpha$. 

Fitting the simulation data to the above function leads to the temperature-dependent values of $\sigma_{\rm p}^{0}$, $A$, and $\alpha$, which are collected in Table \ref{tab1}. These data are mainly based on investigations of ensembles of $N = 3328$ particles; for two temperatures (i.e., for $T = 0.4$ and $T = 0.7$) we have performed complementary investigations for ensembles of $N = 26624$ particles, albeit on a sparser $\dot \gamma$-grid. As expected, the values of the fitting parameters, $\sigma_{\rm p}^{0}$ and $A$ do show a size dependence, which turns out to be logarithmic in nature. 
With these parameters at hand we can draw in panel (a) of Fig.~\ref{fig2a} the solid lines, using the same color code as the one used for the symbols, representing the simulation data for different temperatures. Our data provide evidence that $\sigma_{\rm p}^{0}$ and $A$ decrease with increasing temperature, while the exponent $\alpha (= 0.43)$ remains temperature independent. The good agreement of the fit of the simulation data with the HB type function suggests that $\sigma_{\rm p}$ shows a power-law decay. It seems that the system attains an apparent yield stress, $\sigma_{\rm p}^{0}$, i.e., the value of $\sigma_{\rm p}$ at infinitesimally small shear rates. While we observe in our study a finite value for $\sigma_{\rm p}^0$, recent theories suggest that in the thermodynamic limit this apparent yield stress should disappear \cite{reddy2020nucleation,nath2018existence,sausset2010solids}. 
To investigate this discrepancy further, we explore the dependence of $\sigma_{\rm p}^{0}$ on the system size. Our finite-size analysis suggests that $\sigma_{\rm p}^{0}$ decays logarithmically with increasing system size: the inset in panel (b) of Fig.~\ref{fig2a} shows the variation of $\sigma_{\rm p}^{0}$ as a function of the inverse of the ensemble size for five different values of $N$, namely $N = 1300, 3328, 6500, 26624, 52000$ at $T = 0.7$. The solid back line represents in this panel a fit of these data with the functional form
\begin{eqnarray}
\label{logfit}
\sigma_{\rm p}^{0} = 17.9227 + 0.593078\ln(1/N) .
\end{eqnarray}

The different curves $\sigma_{\rm p} = \sigma_{\rm p}(\dot{\gamma})$ obtained for different temperatures, as shown in panel (a) of Fig.~\ref{fig2a} can be mapped onto one single master curve by scaling the shear rate by a temperature-dependent timescale, $\tau_{\rm c}(T)$, the latter one being defined as $\tau_{\rm c} = \left(A/\sigma^{0}_{\rm p}\right)^{1/\alpha}$; the values of this timescale are listed for different temperatures in Table~\ref{tab1}. Starting our reasoning from Equ. (\ref{hb}) this master curve has thus the form \cite{chaudhuri2012inhomogeneous,shrivastav2020on}:
\begin{eqnarray}
\label{sclhb}
\frac{\sigma_{\rm p} (\dot \gamma)}{\sigma^{0}_{\rm p}} = 1 +
\left(\tau_{c}\dot{\gamma}\right)^{\alpha} .
\end{eqnarray}

\begin{table}[htb]
\caption{Parameters $\sigma_{\rm p}^{0} = \sigma_{\rm p}^{0}(T)$ and $A = A(T)$ as obtained by fitting the simulation data for $\sigma_{\rm p}$ -- see panels (a) and (b) of Fig.~\ref{fig1} -- via the HB type expression, given in Eq. (\ref{hb}); data are listed  for the five different temperatures investigated in this contribution. Data are based on investigations with ensemble sizes as indicated ($N$ is the number of particles). The temperature-dependent timescale $\tau_{\rm c}$ is calculated from $A$ and $\sigma_{\rm p}$ via $\tau_{\rm c} = (A/\sigma_{\rm p}^0) ^{1/\alpha}$.}
\label{tab1}
\vspace{2ex}
\begin{center}
\renewcommand{\arraystretch}{0}
\begin{tabular}{|c|c||c|c|c|}
\hline
$T$ & $N$ & $\sigma^{0}_{\rm p}$ & $A$ & $\tau_{\rm c}$ \strut\\
\hline
0.4  & ~3328 & 16.9348 & 28.467  & 3.3463  \strut \\
0.4  & 26624 & 16.8865 & 24.2684 & 2.3242  \strut \\
\hline
0.5  & ~3328 & 15.9376 & 25.4977 & 2.98266 \strut\\
\hline
0.6  & ~3328 & 14.4088 & 24.7787 & 3.5283  \strut\\
\hline
0.7  & ~3328 & 12.9406 & 24.4795 & 4.4039  \strut  \\
0.7  & 26624 & 11.8012 & 27.967  & 7.4377  \strut \\
\hline
\end{tabular}
\renewcommand{\arraystretch}{1}
\end{center}
\end{table}

In panel (b) of Fig.~\ref{fig2a} we show this master curve onto which the different $\sigma_{\rm p}(\dot \gamma)$-curves can be mapped. This obvious scaling behavior provides evidence that in cluster crystals yielding represents a universal scenario which is independent of temperature and which does not rely on the diffusion of particles. In our considerations the HB model is used to describe the behavior of the steady-state stress as a function of shear rate. We point out that a wide range of viscoelastic materials exhibits a power-law thinning at high shear rates with a temperature-independent exponent \cite{bonn2017yield}. 

\subsubsection{Dyamics under shear}
\label{subsub:shear_msd}

We further analyze the dynamics of point defects under shear for cluster crystals by focusing on the MSD of our particles. As mentioned in Sec.~\ref{subsec:equib}, we consider in the following two different temperatures, i.e.,  $T = 0.4$ and $T = 0.7$. To obtain information on the non-affine displacement of particles under shear we investigate henceforward in the following the $z$-component of the MSD, $\langle (\Delta r_z)^2 \rangle$, since the  $z$-axis represents  the direction of the shear gradient in our setup. 

\begin{figure}
\centerline{\includegraphics[width=0.5\textwidth]{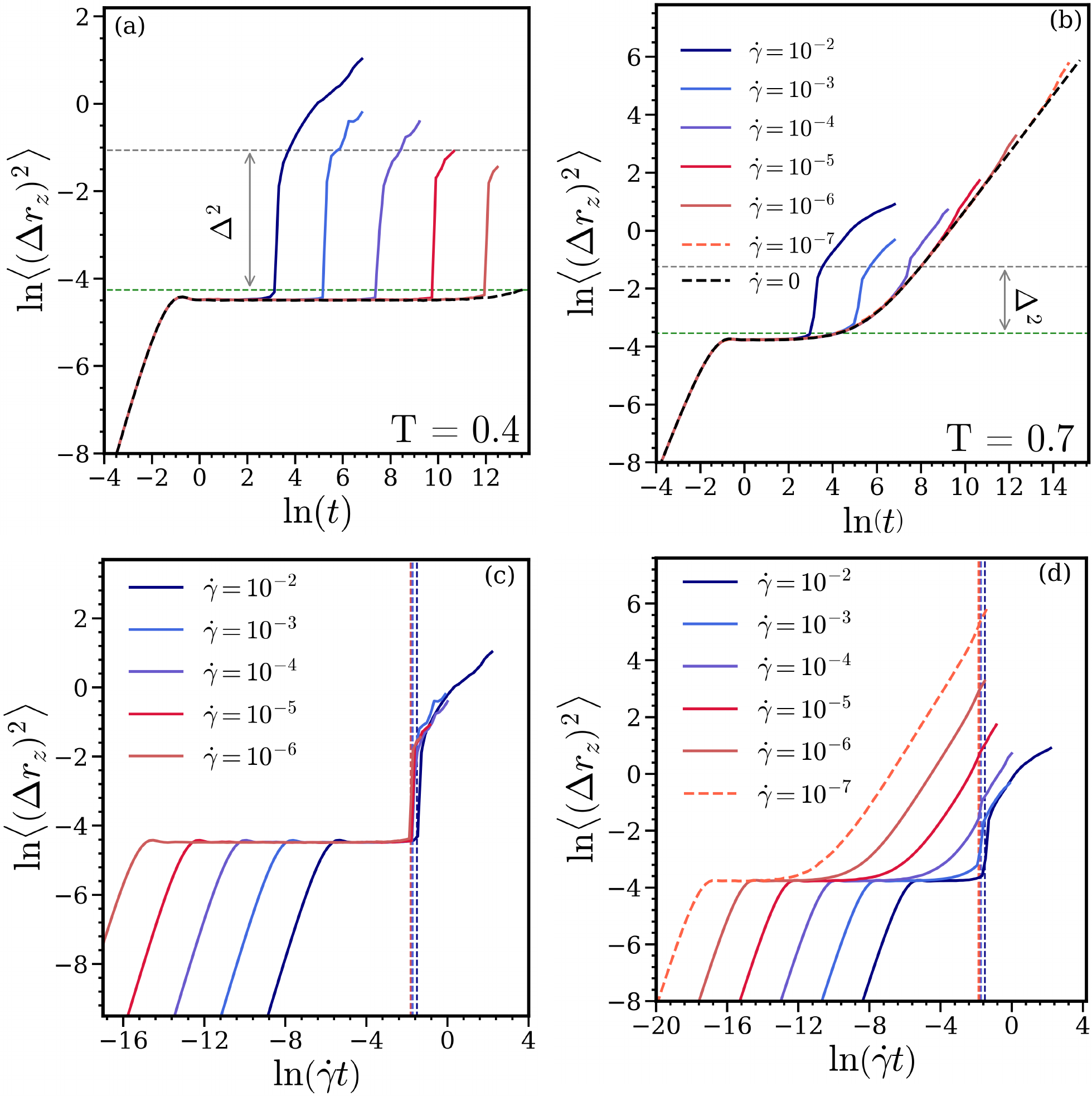}}
\caption{Panels (a) and (b): $z$-component of the MSD, $\langle (\Delta r_z)^2 \rangle$, of a cluster crystal formed by ultrasoft particles in equilibrium and under shear as a function of time $t$ at $T = 0.4$ (a) and $T=0.7$ (b), respectively. The results are shown for six different shear rates, $\dot{\gamma} = 10^{-2}, 10^{-3}, 10^{-4}, 10^{-5}$ and $10^{-6}$ (as labeled in panel (b)) and for the equilibrium state (i.e., for $\dot \gamma =0$). Results are based on ensembles of $N = 26624$ particles except for the MSD of the equilibrium state and for an additional shear rate $\dot{\gamma} = 10^{-7}$ (for $T = 0.7$, in panels (b) and (d)), which are computed from a smaller system with $N = 3328$. These curves are shown by dashed lines in the respective panels. The quantity $\Delta^2$ and the horizontal dashed line are discussed in the text. The green dashed line represents $R_{g}^{2}$ in both the panels. Panels (c) and (d): $z$-component of the MSD, $\langle (\Delta r_z)^2 \rangle$, of a cluster crystal formed by ultrasoft particles in equilibrium and under shear as functions of the strain, $\dot \gamma t$, at $T = 0.4$ (c) and $T = 0.7$ (d), respectively. The results are shown for six different shear rates,  $\dot{\gamma} = 10^{-2}, 10^{-3}, 10^{-4}, 10^{-5}$ and $10^{-6}$ (as labeled in panel (c)) and for the equilibrium state (i.e., for $\dot \gamma =0$). Results are based on ensembles of $N = 26624$ particles. The vertical lines correspond to the maxima in the stress vs. strain curves shown in Fig. \ref{fig2}, i.e., they mark the yield strain for the respective shear rate; note that these lines are drawn in the same colours as the respective MSD curves.}
\label{fig3}
\end{figure}

Panel (a) of Fig.~\ref{fig3} shows $\langle (\Delta r_z)^2 \rangle$ of the particles at $T = 0.4$, considering different values of the shear rates. 
In the absence of shear (i.e., for $\dot \gamma = 0$) we observe two distinctively different types of behaviour of $\langle (\Delta r_z)^2 \rangle$ as a function of time $t$: first, the expected ballistic regime at short times, which then levels off into a plateau (where the MSD attains a value of $\simeq 10^{-2}$). Since this plateau persists for $\dot \gamma = 0$ over the entire duration of the simulation (see the black dashed line in this panel) we conclude that at equilibrium essentially no particle hopping occurs. However, as soon as shear sets in we observe four different regimes in the $\langle (\Delta r_z)^2 \rangle$-curves: beyond the ballistic regime (which coincides throughout with the results obtained for the equilibrium case) a plateau occurs where, again, $\langle (\Delta r_z)^2 \rangle$ assumes a  value of $\simeq 10^{-2}$. Now this plateau is of finite length in time $t$ and its extent decreases with increasing shear rate $\dot \gamma$. This region is then followed by a pronounced superdiffusive regime, where $\langle (\Delta r_z)^2 \rangle$ changes in an essentially discontinuous manner by nearly one order of magnitude: this feature is -- as detailed below -- a consequence of the sudden escape of particles from the clusters. The abrupt change in $\langle (\Delta r_z)^2 \rangle$ -- delimited in panel (a) of Fig. \ref{fig3} by the aforementioned plateau and the dashed horizontal line -- is denoted by $\Delta^2$ (see below). At this point it must be mentioned that the onset of the superdiffusive behaviour coincides with the yielding point of the stress vs. strain curves (see discussion below). In passing we note that such a particular superdiffusive behavior has also been observed in glasses and supercooled liquids under shear, corresponding there to the breaking of neighboring cages of particles \citep{varnik2004study,shrivastav2016yielding}. In our system this superdiffusive regime is followed by a smooth increase of the MSD with time; due to the high numerical costs of the simulations we were not able to collect data of sufficient statistical quality in this time window which would allow us to extract the effective exponent of the MSD and to make thereby more detailed conclusions on the nature of this type of particle transport. Returning to the superdiffusive regime, a more quantitative analysis reveals that $\Delta = (l_{a}/\sqrt{2} - 2R_{g})/2 \simeq 0.58$: in view of the fact that the nearest neighbour distance between two clusters amounts to $l_{a}/\sqrt{2} \simeq 1.41$ and estimating the spatial extent of a cluster via its radius of gyration, $R_{\rm g} \simeq 0.12$ (its value is denoted by green dashed line in panel (a) of Fig.~\ref{fig4}), we conclude that the abrupt change in the MSD must be related to the hopping of the particles from one cluster to a neighbouring one. This interpretation is also supported by two additional facts: (i) the actual value of $\Delta$ is found to be independent of the shear rate; thus we conclude that up to the yielding point particles are either located in clusters or hop from one cluster to another one; (ii) at this low temperature the hopping timescale, $\tau_{\rm h} (\ll \tau_{\rm s})$ assumes an essentially infinite value (see panel (f) of Fig. \ref{fig1}). 

The situation is distinctively different at the higher temperature ($T = 0.7$) and possibly more intriguing; the related data are shown in panel (b) of Fig. \ref{fig3}. Again, the MSD shows in the equilibrium state the trivial ballistic behaviour at short times, followed by a pleateau type region which extends over approximately two orders of magnitude in time; eventually a diffusive process sets in and the MSD grows again with time. Similar as in the low temperature case, the MSD follows under shear initially the related curve of the equilibrium state. Then -- depending on the shear rate -- the curves of the MSD separate from the equilibrium data via a superdiffusive behaviour: the higher the shear rate, the earlier the onset of this regime and the more pronounced the superdiffusivity (both in its in extent and its onset); using the quantity $\Delta$ introduced above, we observe that $\Delta$ decreases as the shear rate is lowered; however for low shear rates (i.e., where $\tau_{\rm s} \gg \tau_{\rm h}$) $\Delta$ is barely noticeable and  the transition between the different diffusive regimes becomes rather smooth. Eventually, for $\dot \gamma = 10^{-7}$ the MSD curves for the sheared and for the equilibrium state are hardly discernable. This indicates that -- although the hopping of particles (or, equivalently the dynamics of the point defects) -- facilitates stress relaxation, the yielding phenomenon is not entirely associated to the diffusion of the particles. We note that cluster crystals are, however,
intermediate between liquid-like and crystalline systems: on one side they share features of periodicity of crystals, on the other side they are characterized by liquid-like diffusion of particles and a disordered intra-cluster disordered structure. Therefore, we expect that the yielding of cluster crystals should be based both on the diffusion of particles and on the deformation of the underlying crystalline skeleton. Another evidence for this hypothesis is provided by the fact that in the stress vs. strain curves their maximum does not disappear at low shear rates where $\tau_{s} \gg \tau_{h}$.

The above mentioned features are in striking contrast to non-Newtonian fluids (such as supercooled liquids \cite{golkia2020flow,zausch2008equilibrium}, ferrofluids \cite{sreekumari2015anisotropy}, polymers \cite{toneian2019controlled}, mixtures of ferrofluids and liquid crystals \cite{shrivastav2020steady}, to name a few) where only diffusion of the particles is responsible for the stress relaxation, and the maximum in the stress vs. strain response disappears at low shear rates, i.e. where $\tau_{s}$ is comparable to structural relaxation times. Also, at these (low) shear rates, the MSD of the particles coincides under shear with the equilibrium MSD \cite{golkia2020flow,zausch2008equilibrium}. In contrast, in our cluster crystals, the MSD of particles under shear deviates from the equilibrium MSD at the yielding point even for the lowest shear rate. This suggests that shear induces a deformation of the underlying FCC skeleton, a feature that we shall investigate in more detail in the following sections.



To round up the discussion we now take an alternative view on the data and discuss the MSDs as functions of the strain, $\dot \gamma t$. As mentioned already briefly above, our data show that the onset of the superdiffusive regime coincides at both temperatures considered with the onset of the yield strain (i.e., with the maxima in the stress vs. strain curves). Panel (c) of Fig.~\ref{fig3} shows the $z$-component of the MSD as a function of strain for $T = 0.4$ and for the shear rates considered in panel (a) of Fig.~\ref{fig3}.  The dashed vertical lines mark the yield strains for this temperature, as extracted from the stress vs. strain curves (see panel (b) of  Fig.~\ref{fig2}). We observe that the yield strain is essentially independent of the shear rate and that the onset of the superdiffusive regime in the MSD occurs essentially when the cluster crystal yields to the stress. The situations is notably different for the higher temperature, with the related data shown in panel (d) of Fig.~\ref{fig3}: again the values of the strain where yielding sets in are essentially independent of the shear rate. However, we observe two distinctively different scenarios for high (i.e., $\dot \gamma \simeq 10^{-2}$ or $10^{-3}$) and low shear rates: in the former case, the superdiffusive behaviour indicates that yielding occurs -- similar as for the low temperature -- via hopping processes from one cluster to the other; in the latter case the smooth variation of the MSDs as functions of strain provide evidence that at lower shear rates yielding is mostly supported by particle diffusion.

In view of the fact that the MSD is a microscopic quantity while the stress-strain response is a macroscopic feature, the above observations suggest that the signatures of yielding at the microscopic scale can be characterized by the displacement of particles from a reference configuration. Such an approach is, for instance, used to identify dynamical heterogeneities at the local scale in glassy systems under shear \cite{shrivastav2016yielding,shrivastav2016heterogeneous}. 
Therefore, to characterize the yielding events at the microscopic level we investigate in the following the structural evolution of the skeleton of the cluster crystal under shear.

\subsubsection{Microstructure under shear}
\label{ms_shear}

As already shown by Coslovich et al. \cite{coslovich2011hopping} particles hop under equilibrium conditions in cluster crystals from one cluster to a neighboring one, while preserving the FCC ordering of the skeleton. In an effort to understand the impact of  shear on this type of particle transport, we start by identifying the skeleton of the cluster crystal by tracing the centers of mass of the clusters and their time evolution under shear. For this purpose, we focus in the following on the higher temperature (i.e., $T = 0.7$): choosing in the following $\dot \gamma = 10^{-3}$ and $\dot \gamma = 10^{-6}$ we obtain insight both to  the high and the low shear rate regimes (see Sec.~\ref{subsec:equib}). 

\begin{figure*}
\centerline{\includegraphics[width=0.9\textwidth]{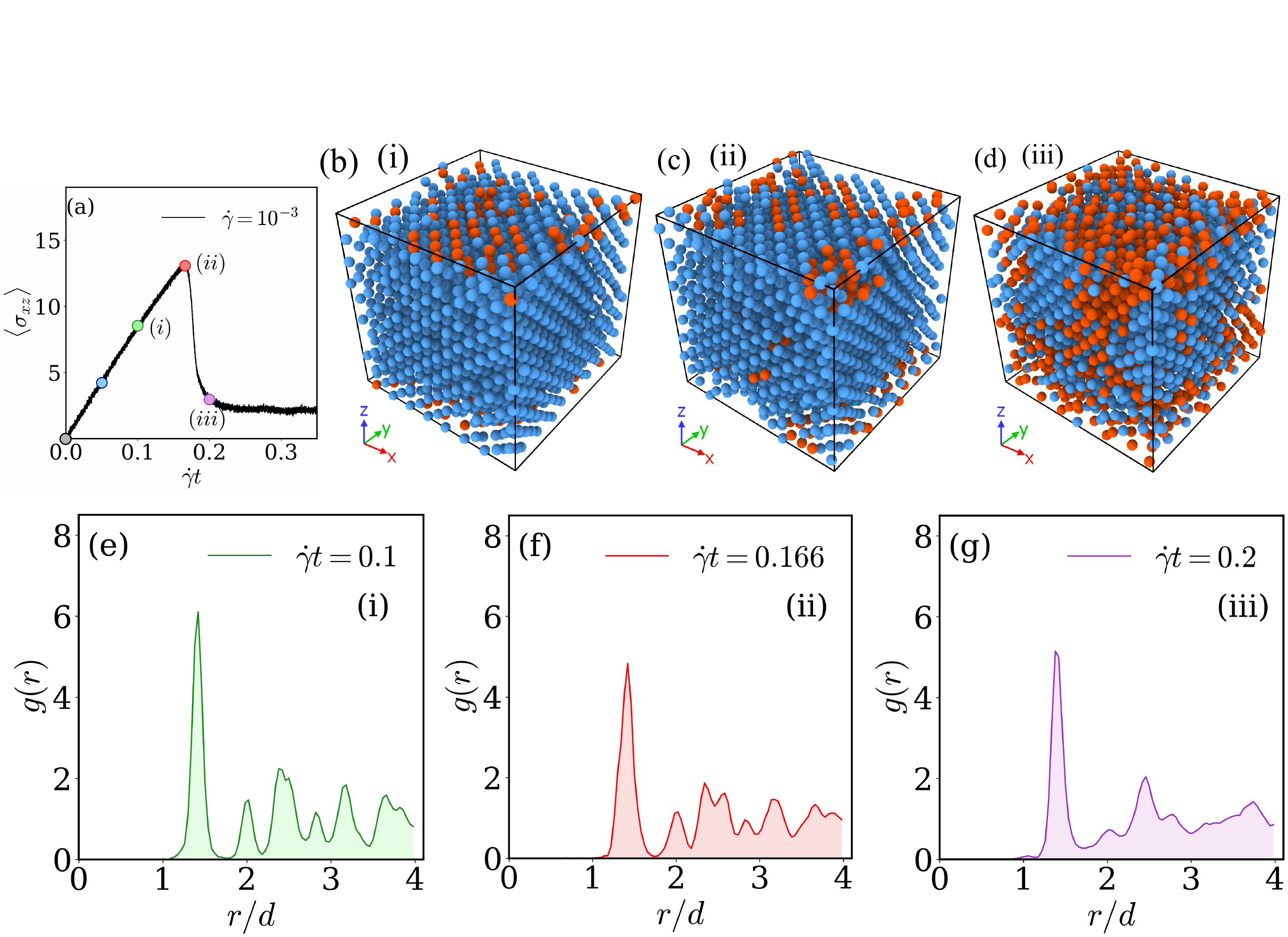}}
\caption{Investigating the skeleton of a cluster crystal under shear -- the high shear rate regime (i.e., $\dot \gamma = 10^{-3}$). Panel (a): stress vs. strain curve for a cluster crystal at $T = 0.7$, exposed to a shear rate of $\dot{\gamma} = 10^{-3}$. Coloured circles along this curve indicate selected values of shear (see text); their numbering refers to the labels of the other panels. Panels (b) -- (d): snapshot of the skeleton (in terms of the centers of mass of the clusters), shown for three different values of strain ((i) -- $\dot \gamma t = 0.1$, (ii) -- $\dot \gamma  t= 0.166$, and (iii) -- $\dot \gamma t = 0.2$): in these snapshots the centers of mass with twelve neighbours are shown in blue, while all other centers of mass are shown in red. Panels (e) -- (g): radial distribution functions, $g(r)$, as functions of the distance, $r$, shown for the very same values of strain as the corresponding snapshots of the skeletons in the panels above. The actual values of strain are shown in the upper right corner of the respective $g(r)$ panels. Results are based on ensembles of 26624 particles.}
\label{fig4}
\end{figure*}

Let us first consider the shear rate $\dot{\gamma} = 10^{-3}$ which represents in our nomenclature the high shear rate regime. We choose one sample of our system and investigate systematically the evolution of the original FCC skeleton as we steadily increase the shear. To this end we have marked along the stress vs. strain curve (shown in panel (a) of Fig. \ref{fig4}) five values for the strain: each of them is marked by a coloured circle and a numeric label: the first one (coloured in grey) corresponds to the equilibrium case, while the strain values marked in blue and green (the latter one with label (i)) are located in the linear response regime; the fourth value of strain (coloured in red and labeled (ii)) is located at the maximum of the stress vs. strain curve, while value (v) (and coloured in purple) has been chosen in the transient regime beyond this maximum. Simulation snapshots corresponding to states (i) to (iii)  are shown in panels (b) to (d) of Fig.~\ref{fig4}; there the centers of mass of the clusters are coloured according to their immediate vicinity: centers of mass with twelve neighbours are shown in blue, while all other points of the skeleton are coloured in red. The respective radial distribution functions (corresponding to these snapshots) are shown in the bottom row of this figure (using the labeling and colouring introduced above). The snapshots and the results for the radial distribution functions suggest that in the linear regime of the stress vs. strain curve the FCC structure of the skeleton is marginally affected by the strain (case (i)): only for a few centers of mass the ideal surrounding of twelve neighbours is disturbed while the peaks in the $g(r)$ are located at the expected positions. However, close to the maximum in the stress vs. strain curve (i.e., case (ii)) the peaks of $g(r)$ broaden and the characteristic pattern of the peak positions is lost; concomitantly a significant number of centers of mass has lost its ideal FCC surrounding. Eventually, in the transient regime (case (iii)) the FCC order of the skeleton is almost completely destroyed: the radial distribution function shows a liquid like behaviour while a few centers of mass have still preserved their ideal number of twelve nearest neighbours. 

\begin{figure*}
\centerline{\includegraphics[width=0.9\textwidth]{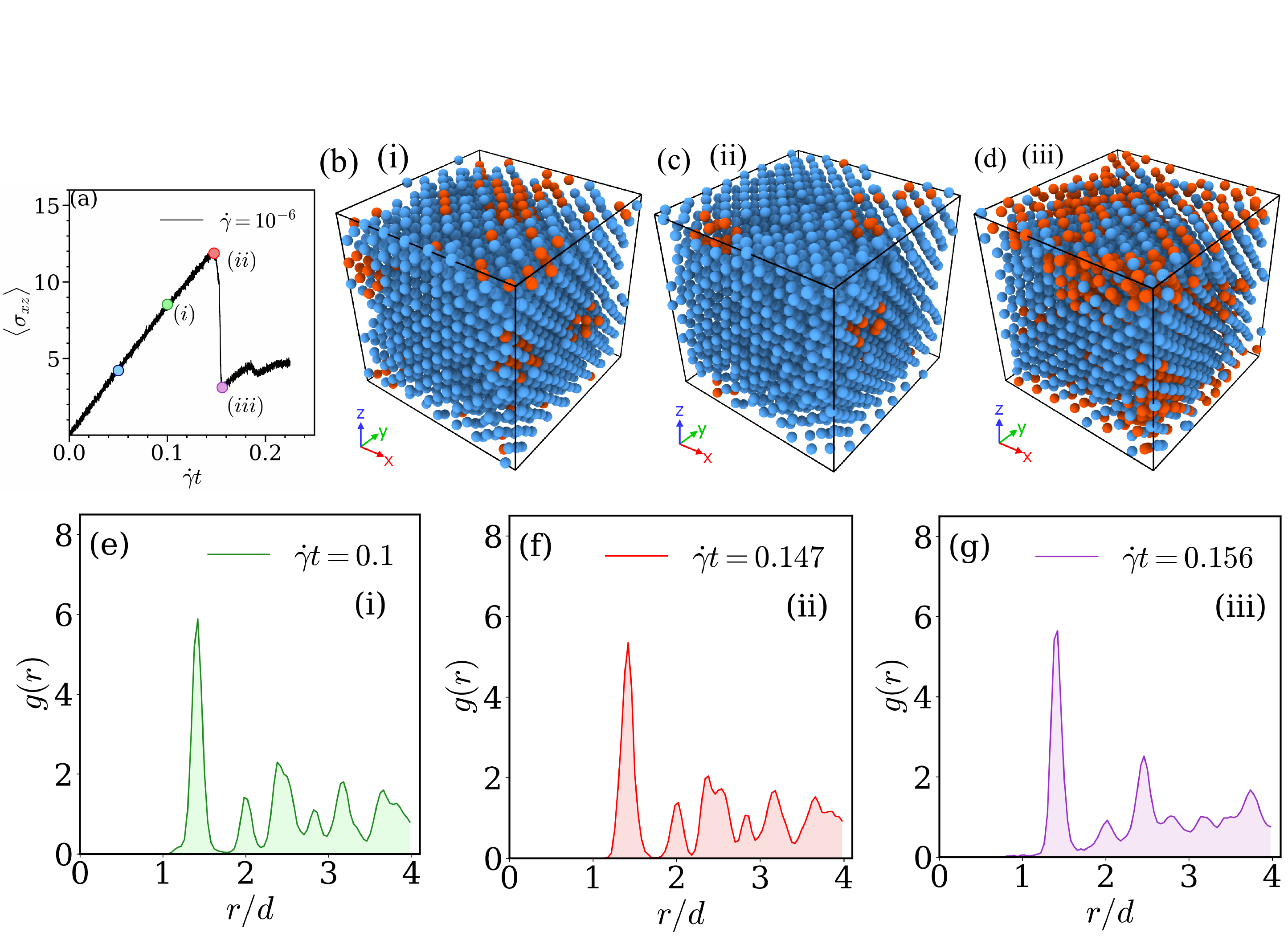}}
\caption{Investigating the skeleton of a cluster crystal under shear -- the low shear rate regime. Investigations have been carried out at $T = 0.7$ and at a shear rate $\dot{\gamma} = 10^{-6}$. Same as Fig. \ref{fig4}. 
}
\label{fig5}
\end{figure*}

For the low shear rate, $\dot{\gamma} = 10^{-6}$, we end up with essentially the same conclusions as for the high shear rate case. Related results are now summarized in the panels of Fig.~\ref{fig5}: panel (a) shows the stress vs. strain curve (defining the five selected values of strain), panels (b) to (d) show the centers of mass of the clusters of a selected simulation snapshot, while panels (d) to (g) show the corresponding radial distribution functions $g(r)$. 

\subsubsection{Characterizing structure at the local scale under shear}
\label{subsubsec:local_structure}

Local crystalline order can be conveniently analysed via the Steinhardt order parameters or local bond order parameters \cite{steinhardt1983bond,lechner2008accurate,rein1996numerical,eslami2018local}. In the following we use these parameters to probe the evolution of the local structure of the skeleton of the cluster crystal under shear. 

To be more specific we calculate bond order parameters $\bar{q}_{6}$ and $\bar{q}_{4}$ which are defined as \cite{lechner2008accurate}:

\begin{eqnarray}
\label{bop}
\bar{q}_{l}(i) = \sqrt{\frac{4\pi}{2l + 1}\sum_{m=-l}^{l}\mid\bar{q}_{lm}(i)\mid^{2}} ,
\end{eqnarray}
with 

\begin{eqnarray}
\label{bop1}
\bar{q}_{lm}(i) = 
\frac{1}{N_{n}(i) + 1}\sum_{k=0}^{N_{n}(i)}q_{lm}(k)
\end{eqnarray}
and
\begin{eqnarray}
\label{bop2}
q_{lm}(i) = \frac{1}{N_{n}(i)}\sum_{j=1}^{N_{n}(i)} Y_{lm}(\hat {\bm r}_{ij}). 
\end{eqnarray}
In relation (\ref{bop1}) the sum includes all $N_n(i)$ neighboring clusters of cluster $i$ and -- via the ($k = 0$)-term -- the cluster $i$ itself. Further, the $Y_{lm}(\hat {\bm r}_{ij})$ are the spherical harmonics with $\hat {\bm r}_{ij}$ being the normalized displacement vector from the center-of-mass of cluster $i$ to the one of cluster $j$. Note that the above definition of local bond order parameters involves an additional averaging over the centers-of-mass of the clusters in the surrounding second shell of cluster $i$. At a given strain, we use the distance of the first minimum in the pair correlation function to determine the neighbors of a center of mass of a cluster. 

\begin{figure}
\centerline{\includegraphics[width=0.5\textwidth]{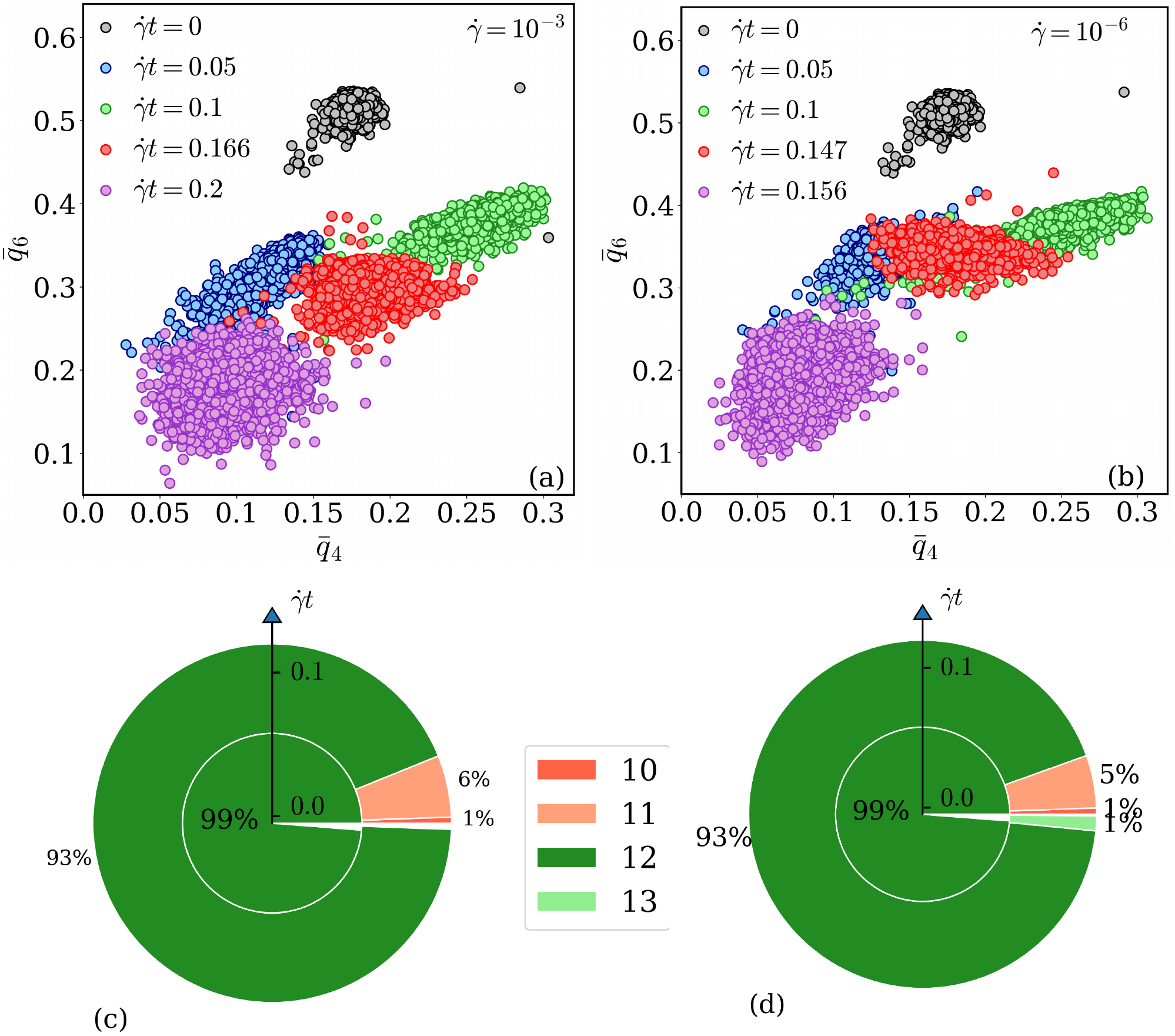}}
\caption{Panel (a): scatter plot of the bond order parameters $\bar{q}_{6}$ and $\bar{q}_{4}$, defined in Eq.~(\ref{bop}) for strain values $\dot{\gamma}t = 0, 0.05, 0.1, 0.166$ and $0.2$. The shear rate is fixed to $\dot{\gamma} = 10^{-3}$, further $T = 0.7$. Panel (b): scatter plot of the bond order parameters $q_{6}$ and $q_{4}$, defined in Eq.~(\ref{bop}) for strain values $\dot{\gamma}t = 0, 0.05, 0.1, 0.147$ and $0.156$. The shear rate is fixed to $\dot{\gamma} = 10^{-6}$, further $T = 0.7$. Panels (c) and (d): nested pie plots showing the fraction of centers of mass of clusters with different numbers of neighbors for shear rates $\dot{\gamma} = 10^{-3}$ (panel (c)) and $\dot{\gamma} = 10^{-6}$ (panel (d)). The direction of the arrow points in both panels in the direction of increasing strain. The color code indicates the number of neighbors. Results are based on ensembles of 26624 particles.}
\label{fig6}
\end{figure}

Panels (a) and (b) of Fig.~\ref{fig6} show scatter plots -- $\bar{q}_{6}$ vs. $\bar{q}_{4}$ -- for a cluster crystal at $T = 0.7$ which is exposed to shear with shear rates $\dot{\gamma} = 10^{-3}$ and $\dot{\gamma} = 10^{-6}$, respectively. Data are plotted for the equilibrium state ($\dot \gamma t = 0$) and for four additional values of strain (as labeled in the panels and using the same colour code); the corresponding states are marked in the stress vs. strain curves shown for the two shear rates in panels (a) of Fig.~\ref{fig4} and Fig.~\ref{fig5}, respectively. 

Starting from the equilibrium state (where we observe an essentially ideal FCC order) we find that both $\bar{q}_{6}$ and $\bar{q}_{4}$ decrease with the increasing strain, indicating a steady decrease in the FCC order. However, in the state close to the yield strain (represented by the red symbols) $\bar{q}_{4}$ increases substantially; note that the growth in $\bar{q_{4}}$ is more pronounced for the lower shear rate. Beyond the yield strain, both bond order parameters assume -- independent of the shear rate -- rather small values (purple symbols), providing evidence that the system is now in a disordered state; this disorder seems to be more pronounced for the higher shear rate. 

The increase of $\bar{q_{4}}$ near the yield point indicates the structural changes that the cluster crystal undergoes at the local level. This feature should also be reflected in the coordination number of the clusters consistent with cubic or square in-planar ordering: the pie plots shown in panels (c) and (d) of Fig.~\ref{fig6} provide estimates for the fraction of cluster centers of mass with different coordination numbers. The arrows in these plots indicate the direction of increasing strain and the color map shows the number of neighbors of a given center of mass. Thus, the inner circle shows the coordination number at $\dot{\gamma}t = 0$, and the outer shell represents the relative values of the coordination numbers for $\dot{\gamma}t = 0.1$, i.e., close to the yield point of the respective shear rates. From these data we conclude that -- irrespective of the shear rate -- close to $93\%$ of the cluster centers of mass have twelve neighbors, indicating that -- although the FCC ordering of these centers decreases with increasing strain -- a large fraction of the centers maintains even close to the yield point the FCC structure. We also learn that the cubic ordering (which should be characterized by eight neighbors around a center of mass) does not occur. Therefore, we conclude that the above mentioned increase in $\bar{q_{4}}$ must be connected to an increase in the in-plane square ordering of the centers of mass.


\begin{figure}
\centerline{\includegraphics[width=0.5\textwidth]{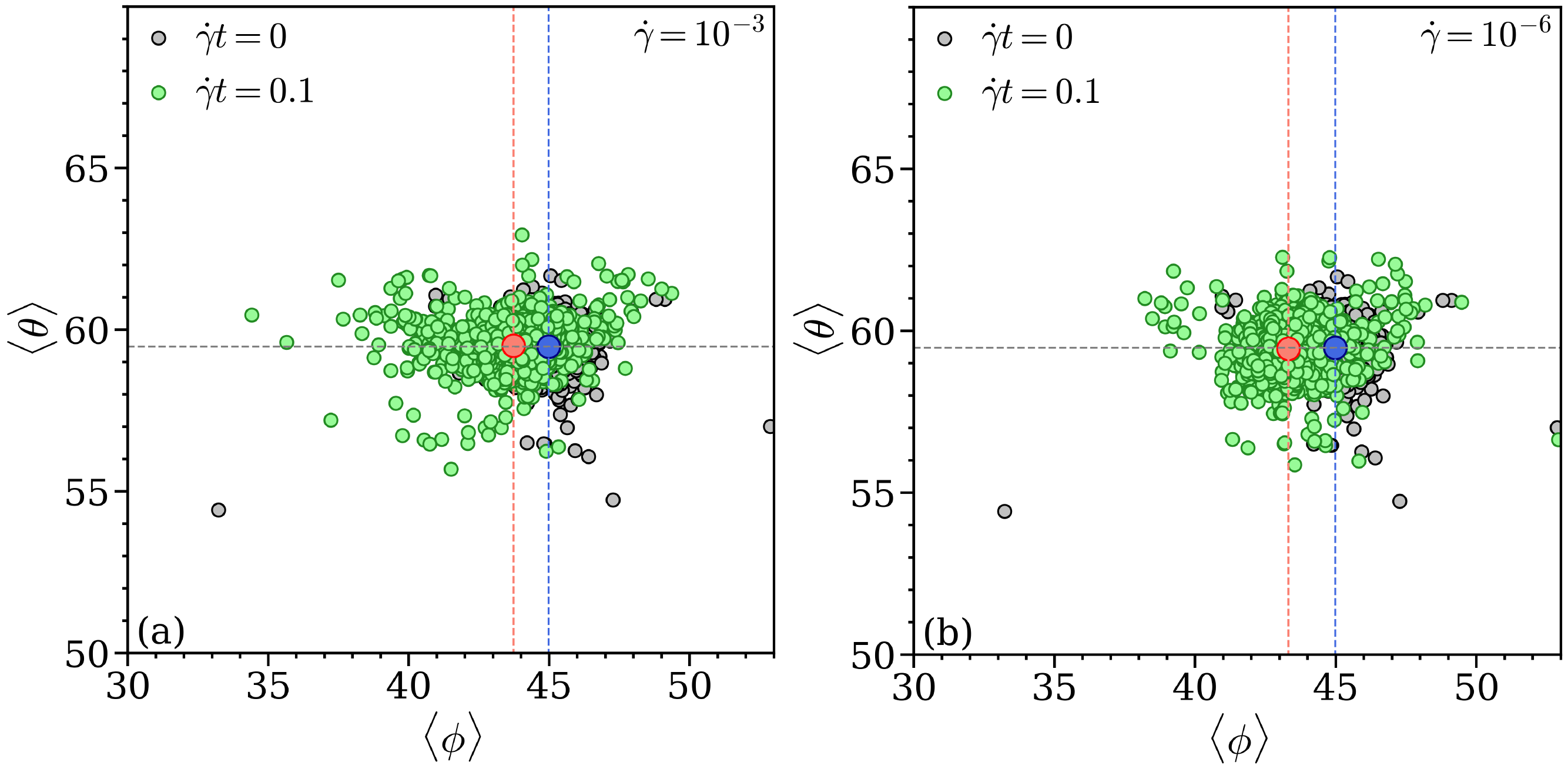}}
\caption{Scatter plots of the average polar $\langle \theta\rangle$ and the average azimuthal $\langle \phi\rangle$ angles in a cluster crystal: black symbols -- equilibrium state, green symbols -- obtained for a strain of $\dot \gamma t = 0.1$; results are based on ensembles of $N = 26624$ particles for different values of the strain (as labeled). Results are shown for shear rates $\dot{\gamma} = 10^{-3}$ (panel (a))  and $\dot{\gamma} = 10^{-6}$ (panel (b)), respectively. The blue and red circles represent in both the panels the center-of-mass of the black and the green data clouds, respectively. The dashed lines mark the position of the red and the blue symbols in the coordinate system}
\label{fig7}
\end{figure}

To investigate this issue further we select among the clusters those which have a $\bar{q_{4}}$-value higher than a threshold value (which we specify below) and calculate the average angle that the center of mass of this cluster encloses with its neighbors. These averages are defined as
\begin{eqnarray}
\label{thph}
\langle \theta(i)\rangle = \sum_{j = 1}^{N_{n}} \theta_{ij}, \;
\langle \phi(i)\rangle = \sum_{j = 1}^{N_{n}} \phi_{ij},
\end{eqnarray}
where, $\theta_{ij}$ and $\phi_{ij}$ are, respectively, the polar and azimuthal angles (with respect to the coordinate system inherent to the simulation cell) between the center of mass of a tagged cluster $i$ and one of its neighbors, carrying the index $j$. The angular brackets denote an average over all the $N_n$ neighboring clusters. It should be noted that in an FCC ordering a cluster has four in-plane neighbors and eight out-of-plane neighbors (i.e., both four in the upper and the lower planes). In the ideal case the values for the in-plane angles are $\phi = 45^{\circ}$ and $\theta = 90^{\circ}$, while for the out-of-plane case the values of the four angles are given by $\phi_1 = 90^{\circ}$, $\theta_1 = 45^{\circ}$, $\phi_2 = 0^{\circ}$, and $\theta_2 = 45^{\circ}$. These values lead to the related average angle $\langle \phi \rangle = 45^{\circ}$ and $\langle \theta \rangle = 60^{\circ}$ for an ideal FCC structure. We note that the angles are restricted throughout to the first quadrant.

For the state where the strain is close to the yield point we choose for the above threshold value $\bar{q}_{4} > 0.2$. The choice for this value of $\bar{q}_{4}$ is motivated by the fact that in equilibrium (where the FCC order is represented in the most ideal manner), $\bar{q}_{4}$ reaches values up to $\bar q_4 \sim 0.2$ (see panels (a) and (b) in Fig. \ref{fig6}). We note that we have chosen for the equilibrium state clusters with $\bar{q}_{4} > 0.12$ as this value is closer to the lower bound of the scatter plot $\bar q_4$ vs. $\bar q_6$ of the equilibrium state; however, this choice has no specific relevance for the following discussion. Panels (a) and (b) of Fig.~\ref{fig7} show the related scatter plots $\langle \theta \rangle$ vs. $\langle \phi \rangle$ in equilibrium ($\dot \gamma t = 0$) and close to the yield point ($\dot \gamma t = 0.1$) for shear rates $\dot{\gamma} = 10^{-3}$ and $10^{-6}$, respectively; in addition the centers-of-mass of these data clouds are highlighted. From the data compiled in Fig.~\ref{fig7} we can make the following conclusions: while we obtain for the equilibrium state the expected values, i.e., $\langle \phi \rangle = 45^{\circ}$ and $\langle \theta \rangle = 60^{\circ}$, the value for $\langle \phi \rangle$ slightly decreases by $\sim 3^{\circ}$) close to the yield point (irrespective of the shear rate); in contrast, the average polar angle, $\langle \theta \rangle$, remains essentially unaffected by the shear. These findings are a clear indication of a planar shift of clusters along the $x$-direction. 

\begin{figure}
\centerline{\includegraphics[width=0.5\textwidth]{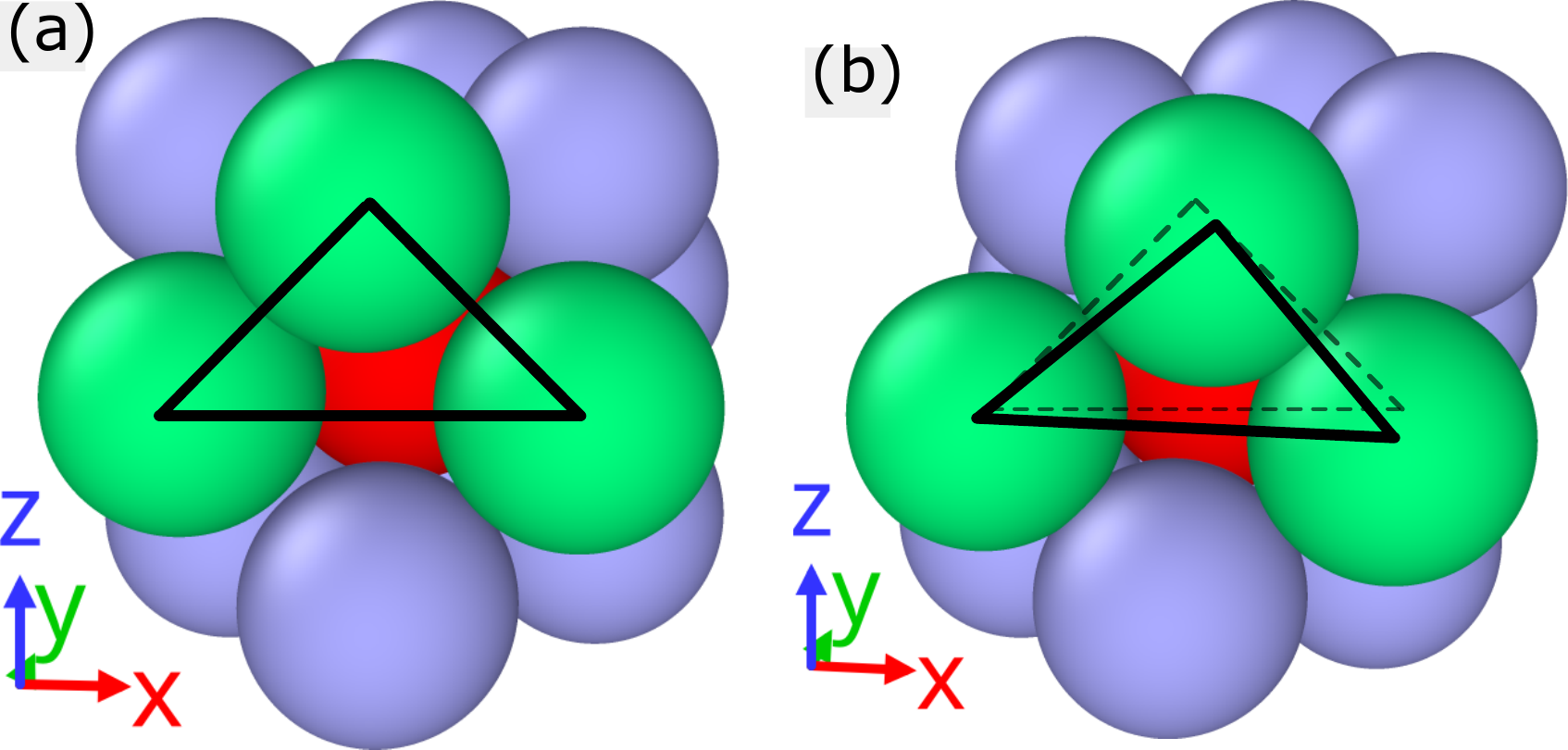}}
\caption{Panel (a): immediate neighbourhood of a tagged cluster (red) in the equilibrium state at $T = 0.7$. Neighbouring clusters are shown in blue and green: the three neighbors located in the  forefront $(x, z)$-plane are marked in green and are connected by solid black lines. Panel (b): immediate neighbourhood of a tagged cluster (red) with $\bar{q}_{4} = 0.2234$, as obtained under shear at a strain $\dot{\gamma}t = 0.18$ (the shear rate is $\dot{\gamma} = 10^{-6}$). The color coding of the spheres is same as the one in panel (a). The solid black lines connect the green spheres while the dashed black lines represent the related configuration of green spheres in the equilibrium state.}
\label{fig8}
\end{figure}

\begin{figure}
\centerline{\includegraphics[width=0.5\textwidth]{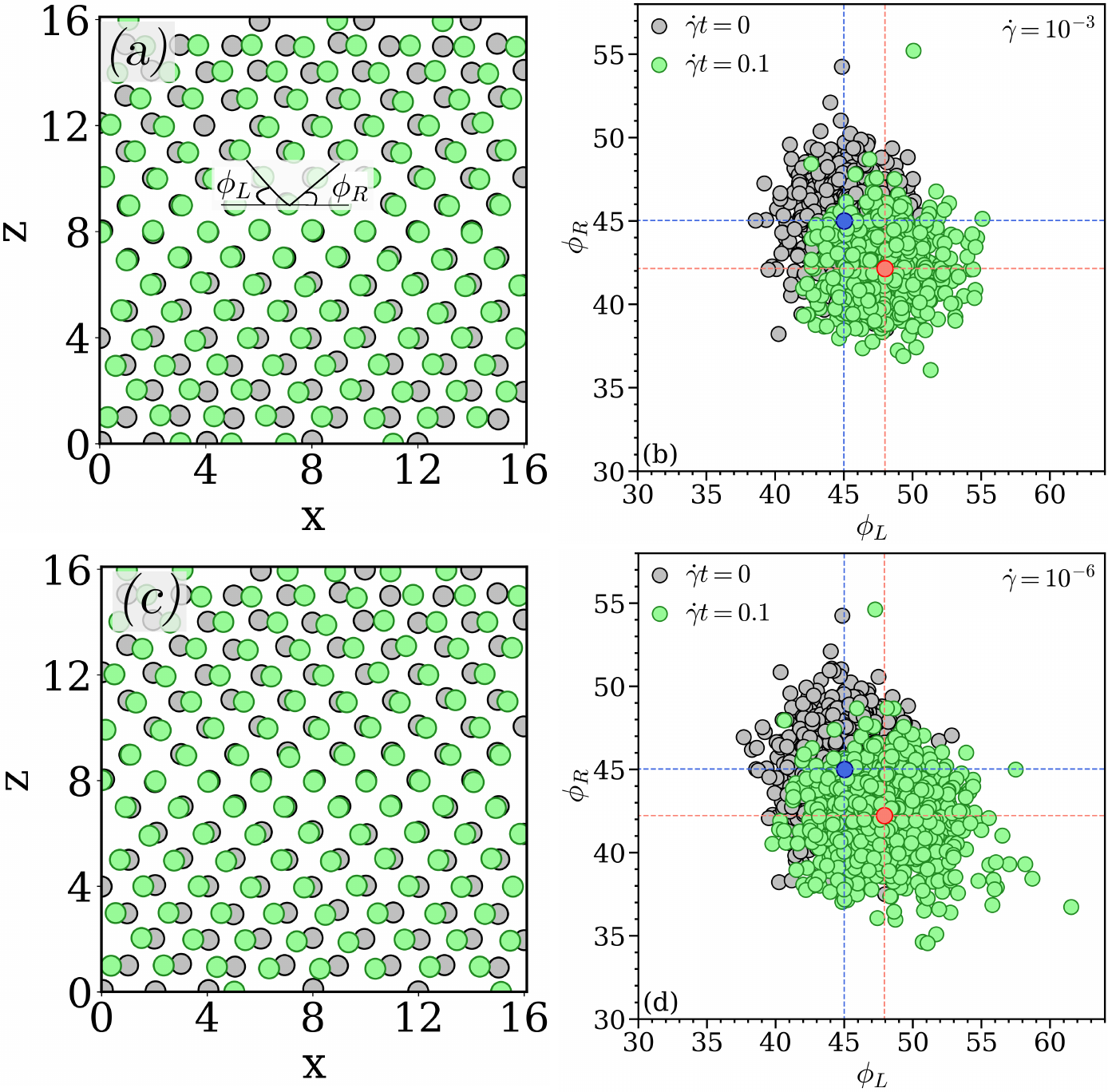}}
\caption{Panel (a): the COMs in the $x-z$ planes in the equilibrium state (filled black circles) and at strain $\dot{\gamma}t = 0.1$ (filled green circles) for shear rate $\dot{\gamma} = 10^{-3}$. The solid black lines represent planar angels $\phi_{L}$ and $\phi_{R}$. Panel(b): scatter plot of the planar angles $\phi_{L}$ and $\phi_{R}$ for shear rate $\dot{\gamma} = 10^{-3}$ at strains, $\dot{\gamma}t = 0$ and $0.1$. Panel (c): the COMs in the $x-z$ planes, shown in Fig.~\ref{8}(d), in the equilibrium state (filled black circles) and at strain $\dot{\gamma}t = 0.1$ (filled green circles) for shear rate $\dot{\gamma} = 10^{-6}$. Panel(d): scatter plot of the planar angles $\phi_{L}$ and $\phi_{R}$ for shear rate $\dot{\gamma} = 10^{-6}$ at strains, $\dot{\gamma}t = 0$ and $0.1$. The blue and red circles in panels (b) and (d) represent the centre of mass of the scatter-cluster at strains $\dot{\gamma}t = 0$ and $0.1$. The blue and red dashed lines show the angles $\phi_{L}$ and $\phi_{R}$ of the COM of the scatter-clusters.}
\label{fig9}
\end{figure}

To better understand this issue we choose one cluster with large $\bar{q}_{4}$-value and analyze the arrangement of its neighboring clusters. Panels (a) and (b) of Fig.~\ref{fig8} show the positions of neighboring clusters  around a tagged cluster equilibrium  and for a cluster (with $\bar{q}_{4} = 0.2234$) at $\dot{\gamma}t = 0.15$ (obtained for a  shear rate $\dot{\gamma} = 10^{-6}$), respectively. In both cases, the central cluster is shown in red while its neighboring clusters are coloured in green and blue: the three neighbors located on the foreground $(x, z)$-plane are highlighted in green. In the equilibrium state, the green COMs are arranged at the edges of an equilateral triangle; in contrast, at the yield point, one can notice a substantial shift of green clusters along the positive $x$-direction (see panel (b) of Fig.~\ref{fig8}): in this panel the dashed triangle connects the positions of the cluster in equilibrium. This finding suggests that a shift of the clusters takes place in the ($x, z)$-plane along the $x$-direction. Experiments and computer simulations on soft colloidal crystals in two dimensions reveal that stress relaxation involves a highly cooperative movement of particles in the system \cite{van2014highly}. We, therefore, expect a cooperative motion of clusters in the $x-z$ plane along the shear direction. In an effort to further analyse the cooperative movement of the clusters in the shear plane along the shear direction, we choose $x-z$ planes at a given strain and calculate angles that a COM makes with its neighbors in the upper layer within the same plane. Notice that in our case, the $z$-axis denotes the direction of the gradient. Hence, under shear, different $x-y$ planes are moving along the $x$-direction, and a $x-z$ plane would correspond to a vertical cross-section of all the moving layers for a given $y$. Two such representative $x-z$ planes at a central $y$ position, are shown in Fig.\ref{fig9} (a) and (c) for the two considered shear rates. The angles $\phi_{L}^{i}$ and $\phi_{R}^{i}$ that $i$-th COM makes with its left and right neighbor in the upper layer are marked in Fig.~\ref{fig9} (a). For all the COMs we calculate these planar angles and plot in Fig.~\ref{fig9} (b) and (d) for shear rates, $\dot{\gamma} = 10^{-3}$ and $10^{-6}$ respectively. Clearly, in the equilibrium state, both the angles remain close to $45^{\circ}$ as indicated by the center of mass (blue circle in Panel (b)) of the scatter-cluster. On the other hand, near the yield point at $\dot{\gamma}t = 0.1$, $\phi_{R}$ decreases by $\sim 3^{\circ}$ and $\phi_{L}$ increases by the same amount, indicating the cooperative planar movement of COMs in different layers along the $x$-direction. 

\subsubsection{Comparison with the soft-sphere FCC crystal under shear}
We finally compare the shear response of cluster crystals and soft-sphere FCC system. The soft-sphere system consist of $N_{SS} = 4000$ particles interacting via WCA potential defined in Eq.~(\ref{wca}) at temperature, $T = 0.01$. In the equilibrium, this system has an FCC order and contains defects due to the finite-size of the system and periodic boundary conditions \cite{swope1992thermodynamics}. However, the defect concentration is small. The works of Bennett et al. \cite{bennett1971studies} and Pronk et al. \cite{pronk2001point} provide an estimate of the point defect concentration in hard-sphere FCC crystals, which is of the order of $10^{-8}$. Although our SS system has a relatively soft core due to $1/r^{12}$ behavior, we do not expect much difference in the defect concentration from the pure hard-sphere system. Moreover, the low temperature ensures that the defect concentration in the SS systems remains negligible compared to the CCs in equilibrium. 

\begin{figure*}[htb]
\centerline{\includegraphics[width=\textwidth]{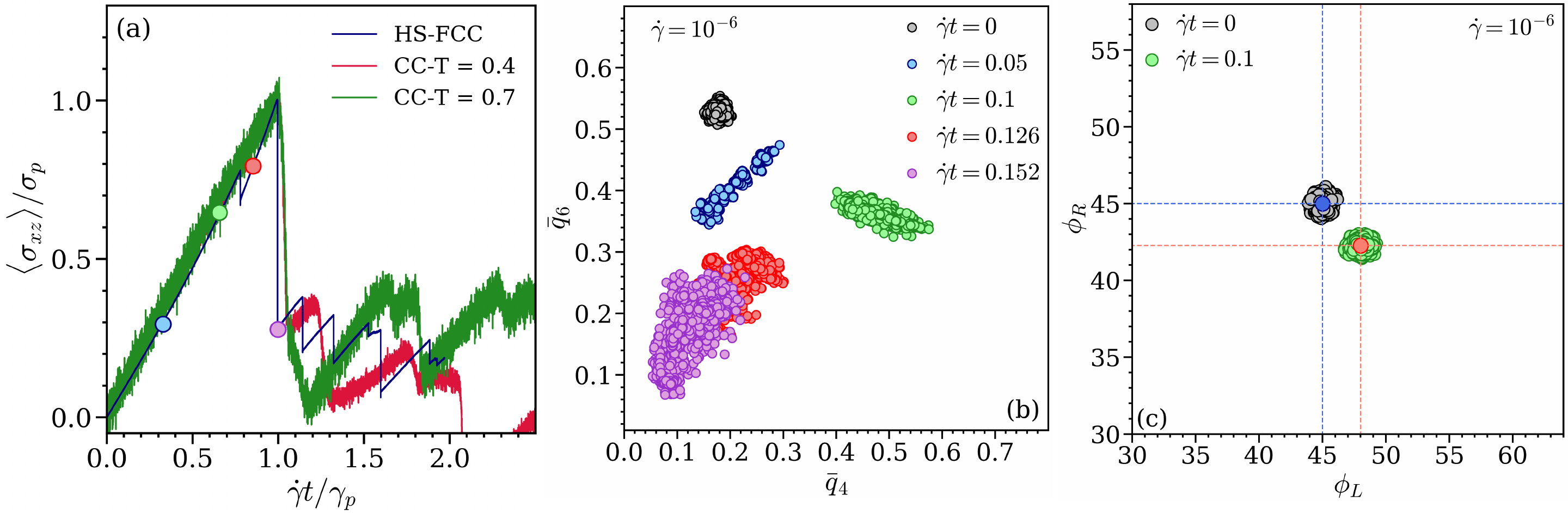}}
\caption{Panel (a): Evolution of scaled stress, $\sigma_{xz}/\sigma_{\rm p}$, as a function of scaled strain, $\dot{\gamma}t/\gamma_{\rm p}$, for a hard sphere FCC (HS-FCC) system and cluster crystals at $T = 0.4$ and $0.7$. All the systems are sheared with the rate $\dot{\gamma} = 10^{-6}$. Panel (b): scatter plot of averaged local bond order parameters, $\bar{q}_{6}$ and $\bar{q}_{4}$ for the HS-FCC system sheared with $\dot{\gamma} = 10^{-6}$ at strains, $\dot{\gamma}t = 0, 0.05, 0.1, 0.13, 0.152$. Panel (c): scatter plot of the planar angles, $\phi_{L}$ and $\phi_{R}$, for the HS-FCC system sheared with $\dot{\gamma} = 10^{-6}$ at strains, $\dot{\gamma}t = 0, 0.152$. The blue and red circles in panel (c) represent the centre of mass of the scatter-cluster at strains $\dot{\gamma}t = 0$ and $0.1$. The blue and red dashed lines show the angles $\phi_{L}$ and $\phi_{R}$ of the COM of the scatter-clusters.} 
\label{fig10}
\end{figure*}

We shear this system with the shear rate $\dot{\gamma} = 10^{-6}$ following the same shearing protocol as for CCs. The stress vs. strain responses of the HS system, together with the CCs are compared in Fig.~\ref{fig10}(a). We have scaled the stress by its peak value and strain by the yield strain. One can observe a similar response of the two systems until the yield point.

Furthermore, the local bond order parameters, $\bar{q}_{6}$ and $\bar{q}_{4}$, shown in Fig.~\ref{fig10}(b), display the trend similar to CCs, cf. Fig.~\ref{fig6}(a) and (b). The only difference here is that the $\bar{q}_{4}$ at $\dot{\gamma}t = 0.1$ (close to the stress maximum) is much higher than the CC system. As expected, this is linked to the cooperative planer movement of particles which is tracked via the planer angles $\phi_{L}$ and $\phi_{R}$. Fig.~\ref{fig10}(c) demonstrates the angles $\phi_{L}$ and $\phi_{R}$ in the equilibrium state and at $\dot{\gamma}t = 0.1$, close to the yield point. Both the angles remain distributed around $45^\circ$ in the equilibrium state as indicated by the blue dashed lines passing through the center of mass of the scatter-cluster (shown by blue circle). This is in contrast to the situation close to the yield point, at $\dot{\gamma}t = 0.1$ where $\phi_{L}$ increases by $\sim 3^{\circ}$ and $\phi_{R}$ decreases by the same amount, which is consistent with the observations for the CCs, cf. Fig.~\ref{fig9} (b) and (d). Thus, we find that the yielding behavior of the cluster crystals is very similar to the FCC crystals with a difference that certain behaviors such as the increase in the $\bar{q}_{4}$ and changes in the planar angles under shear are less pronounced, which may be due to the diffusion of the point defects. This indicates that diffusion of particles (or equivalently, point defects) should strongly affect the transient behavior of the CCs under shear and facilitate shear-induced flow in these systems.
\section{Summary and outlook}
\label{sec:conclusion_outlook}
To summarize, we have studied the yielding of cluster crystals, which represents a model for defect-rich crystals. Our MD simulations reveal that the yielding of cluster crystals depends on the deformation of the underlying FCC structure, and diffusion of individual particles does not affect the yielding scenario. We have selected CC samples at two different temperatures, a higher temperature at which diffusion of particles is more pronounced and a lower temperature where particles remain localized in their respective clusters. The main focus of the work is to understand the role of the diffusion timescale of particles on the yielding behavior under shear. We find that at both temperatures the stress-strain curve at different shear rates has a maximum. The height of this maximum depends on the shear rate, which can be captured by the Herschel-Bulkley type of functional form. Furthermore, the stress-maximum vs. shear rate curves for different temperatures can be scaled onto a master curve, which indicates that the macroscopic yielding behavior of cluster crystals remains independent of the temperature. 

Interestingly, at the microscopic level, by investigating MSDs under shear, we find that for large shear rates yielding is always associated with a superdiffusive behavior of MSDs. At low shear rates, on the other hand, faster diffusion of particles is observed. 
We further identify the center of mass of clusters and demonstrate the deformation of cluster crystals involves the yielding of the underlying FCC structure. The bond order parameters ($q_{6}, q_{4}$) show that the FCC ordering of the center of mass of clusters reduces under shear at all shear rates. The $q_{4}$ order, however, slightly increases close to the yield point. This increase is higher at low shear rates. We find that the increase in the $q_{4}$ order is associated with the cooperative movement of particles in different layers along the direction of shear.

We find that the yielding scenario in cluster crystals is similar to the soft-sphere FCC crystal, which is in contrast to the yielding in amorphous solids where diffusion of particles is the primary process for stress relaxation. It has been observed that in the supercooled liquid state where diffusion time scales are accessible, the dynamics at low shear rates (large shear-induced timescales) remains similar to the equilibrium dynamics \cite{zausch2008equilibrium}. Cluster crystals are peculiar systems where particles, keep diffusing yet maintain the FCC ordering of the system. The diffusion of particles at long timescales is the result of hopping from a lattice site to a neighboring site, which involves a characteristic length scale. For a crystalline system, this length scale is equal to the distance of the nearest neighbor lattice site ($l_{p}/\sqrt{2}$ for FCC). Plastic deformation of crystal results in the modification of this characteristic length scale. This, in turn, appears in the fast diffusion of particles even though the shear-induced timescales are much larger than the equilibrium diffusion timescales. 

In the equilibrium, it has been observed that the longtime diffusive dynamics of cluster crystals involve the power-law distribution of hopping lengths \cite{coslovich2011hopping}. It will be interesting to check (i) if the longtime, i.e., steady-state, dynamics of CCs becomes diffusive and (ii) the behavior of the hopping lengths of particles. Earlier non-equilibrium computer simulations of CCs suggest the formation of string-like structures in the steady-state at large shear rates \cite{nikoubashman2011cluster}. In this case, one expects trapping of particles in strings which may result in an anisotropic dynamics where particles diffuse normally along the string while remain subdiffusive in the direction perpendicular to the string.

We note that in cluster crystals, it is challenging to assign a unique label to the center of mass of the clusters as clusters keep reforming with increasing strain. Therefore, tracking the time evolution of the center of mass or calculating two-time quantities such as displacement of particles from a reference time is challenging. In this work, we do not attempt to calculate such quantities for the center of mass of clusters. We instead compare our results at two or more strain points without matching the labels of the center of mass of the clusters.  

Our results suggest that {\it the diffusion of particles is not the primary mechanism of stress relaxation} in cluster crystals. It is the deformation of the underlying FCC structure, which is responsible for the stress relaxation. We expect that these results will remain valid for defect-rich crystals, which would, then, correspond to the fact that the topological defects play an essential role in the yielding of such crystals. Therefore, it will be interesting to compare our results with the predictions of the recently proposed microscopic theory for the deformation of defect-free crystals \cite{nath2018existence,reddy2020nucleation}. Our results demonstrate that the diffusion of point-defects strongly affects the yield stress and transient behavior of crystals under shear. However, a detailed investigation of the transient response of cluster crystals is required. Furthermore, the characterization of topological defects in the skeleton is needed to get more insight into the mechanism of yielding in these systems. {\it Walk in these directions is on the way.} 

\section{Acknowledgments}
We gratefully acknowledge financial support by the Austrian Science
Foundation (FWF) under Proj. No. I3846. The computational results presented have been achieved using the Vienna Scientific Cluster (VSC). GPS would like to thank J\"{u}rgen Horbach and Saswati Ganguly for fruitful discussions.

\bibliography{ref} 

\bibliographystyle{unsrt}

\end{document}